\documentclass[11pt]{article}

\newcommand{\be}{\begin{equation}}
\newcommand{\ee}{\end{equation}}
\usepackage{amsmath}
\usepackage{amssymb}
\usepackage{latexsym}
\usepackage{epsfig}
\usepackage{color}
\usepackage[makeroom]{cancel}

\newcommand{\half}{{\textstyle {1\over 2}}}

\newcommand{\RRR}{{\hbox{\rm R\kern-2.35mm R}}}

\def\ZZZ{{\hbox{ Z\kern-1.6mm Z}}}

\newcommand{\sectiono}[1]{\section{#1}\setcounter{equation}{0}}

\def\oonoo#1#2#3{\vbox{\ialign{##\crcr
	\hfil\hfil\hfil{$#3{#1}$}\hfil\crcr\noalign{\kern1pt\nointerlineskip}
	$#3{#2}$\crcr}}}
\def\oon#1#2{\mathchoice{\oonoo{#1}{#2}{\displaystyle}}
	{\oonoo{#1}{#2}{\textstyle}}{\oonoo{#1}{#2}{\scriptstyle}}
	{\oonoo{#1}{#2}{\scriptscriptstyle}}}
\def\dt#1{\oon{\hbox{\bf .}}{#1}}  
\def\ddt#1{\oon{\hbox{\bf .\kern-1pt.}}#1}    % À À   (see below)
\def\slap#1#2{\setbox0=\hbox{$#1{#2}$}
	#2\kern-\wd0{\hfuzz=1pt\hbox to\wd0{\hfil$#1{/}$\hfil}}}
\def\sla#1{\mathpalette\slap{#1}}                % slash: see Ö   below
\def\bop#1{\setbox0=\hbox{$#1M$}\mkern1.5mu
	\lower.02\ht0\vbox{\hrule height0pt depth.06\ht0
	\hbox{\vrule width.06\ht0 height.9\ht0 \kern.9\ht0
	\vrule width.06\ht0}\hrule height.06\ht0}\mkern1.5mu}
\def\bo{{\mathpalette\bop{}}}                        % box: see õ below
\mathcode`\*="702A                  % * now always complex conjugate
\def\in{\relax\ifmmode\mathchar"3232\else{\refit in\/}\fi} % ã below 
\def\half{{\textstyle{1\over{\raise.1ex\hbox{$\scriptstyle{2}$}}}}}

\def\Gamma{\mathchar"0100}
\def\Delta{\mathchar"0101}
\def\Theta{\mathchar"0102}
\def\Lambda{\mathchar"0103}
\def\Xi{\mathchar"0104}
\def\Pi{\mathchar"0105}
\def\Sigma{\mathchar"0106}
\def\Upsilon{\mathchar"0107}
\def\Phi{\mathchar"0108}
\def\Psi{\mathchar"0109}
\def\Omega{\mathchar"010A}

\catcode128=13 \def €{\"A}                 % Option-u A
\catcode129=13 \def {\AA}                 % Option-A
\catcode130=13 \def '{\c}           	   % Option-C (cedilla)
\catcode131=13 \def ƒ{\'E}                   % Option-e E
\catcode132=13 \def "{\~N}                   % Option-n N
\catcode133=13 \def …{\"O}                 % Option-u O
\catcode134=13 \def †{\"U}                  % Option-u U
\catcode135=13 \def ‡{\'a}                  % Option-e a
\catcode136=13 \def ˆ{\`a}                   % Option-`  a
\catcode137=13 \def ‰{\^a}                 % Option-i a
\catcode138=13 \def Š{\"a}                 % Option-u a
\catcode139=13 \def ‹{\~a}                   % Option-n a
\catcode140=13 \def Œ{\alpha}            % Option-a
\catcode141=13 \def {\chi}                % Option-c
\catcode142=13 \def Ž{\'e}                   % Option-e e
\catcode143=13 \def {\`e}                    % Option-`  e
\catcode144=13 \def {\^e}                  % Option-i e
\catcode145=13 \def '{\"e}                % Option-u e
\catcode146=13 \def '{\'\i}                 % Option-e i
\catcode147=13 \def "{\`\i}                  % Option-`  i
\catcode148=13 \def "{\^\i}                % Option-i i
\catcode149=13 \def •{\"\i}                % Option-u i
\catcode150=13 \def –{\~n}                  % Option-n n
\catcode151=13 \def —{\'o}                 % Option-e o
\catcode152=13 \def ˜{\`o}                  % Option-`  o
\catcode153=13 \def ™{\^o}                % Option-i o
\catcode154=13 \def š{\"o}                 % Option-u o
\catcode155=13 \def ›{\~o}                  % Option-n o
\catcode156=13 \def œ{\'u}                  % Option-e u
\catcode157=13 \def {\`u}                  % Option-`  u
\catcode158=13 \def ž{\^u}                % Option-i u
\catcode159=13 \def Ÿ{\"u}                % Option-u u
\catcode160=13 \def  {\tau}               % Option-t
\catcode162=13 \def ¢{\oplus}           % Option-4
\catcode163=13 \def £{\relax\ifmmode\to\else\itemize\fi} % Option-3
\catcode164=13 \def ¤{\subset}	  % Option-6
\catcode165=13 \def ¥{\infty}           % Option-8
\catcode166=13 \def ¦{\mp}                % Option-7
\catcode167=13 \def §{\sigma}           % Option-s
\catcode168=13 \def ¨{\rho}               % Option-r
\catcode169=13 \def ©{\gamma}         % Option-g
\catcode170=13 \def ª{\leftrightarrow} % Option-2 ; Option-E (acute) :
\catcode171=13 \def «{\relax\ifmmode\expandafter\acute\else\expandafter\'\fi}
\catcode172=13 \def ¬{\relax\ifmmode\expandafter\ddt\else\expandafter\"\fi}
\catcode173=13 \def ­{\equiv}            % Option-= ; ^ Option-U (umlaudt)
\catcode174=13 \def ®{\approx}          % Option-"
\catcode175=13 \def ¯{\Omega}          % Option-O
\catcode176=13 \def °{\otimes}          % Option-5
\catcode177=13 \def ±{\ne}                 % Option-+
\catcode178=13 \def ²{\le}                   % Option-,
\catcode179=13 \def ³{\ge}                  % Option-.
\catcode180=13 \def ´{\upsilon}          % Option-y
\catcode181=13 \def µ{\mu}                % Option-m
\catcode182=13 \def ¶{\delta}             % Option-d
\catcode183=13 \def ·{\epsilon}          % Option-w
\catcode184=13 \def ¸{\Pi}                  % Option-P
\catcode185=13 \def ¹{\pi}                  % Option-p
\catcode186=13 \def º{\beta}               % Option-b
\catcode187=13 \def »{\partial}           % Option-9
\catcode188=13 \def ¼{\nobreak\ }       % Option-0
\catcode189=13 \def ½{\zeta}               % Option-z
\catcode190=13 \def ¾{\sim}                 % Option-'
\catcode191=13 \def ¿{\omega}           % Option-o
\catcode192=13 \def À{\dt}                     % Option-?
\catcode193=13 \def Á{\gets}                % Option-1
\catcode194=13 \def Â{\lambda}           % Option-l
\catcode195=13 \def Ã{\nu}                   % Option-v
\catcode196=13 \def Ä{\phi}                  % Option-f
\catcode197=13 \def Å{\xi}                     % Option-x
\catcode198=13 \def Æ{\psi}                  % Option-j
\catcode199=13 \def Ç{\int}                    % Option-\
\catcode200=13 \def È{\oint}                 % Option-|
\catcode201=13 \def É{\relax\ifmmode\cdot\else\vol\fi}    % Option-;
\catcode202=13 \def Ê{\relax\ifmmode\,\else\thinspace\fi}
\catcode203=13 \def Ë{\`A}                      % Option-`  A ; ^ Option-space
\catcode204=13 \def Ì{\~A}                      % Option-n A
\catcode205=13 \def Í{\~O}                      % Option-n O
\catcode206=13 \def Î{\Theta}              % Option-Q
\catcode207=13 \def Ï{\theta}               % Option-q; Option-- :
\catcode208=13 \def Ð{\relax\ifmmode\expandafter\bar\else\expandafter\=\fi}
\catcode209=13 \def Ñ{\overline}             % Option-_
\catcode210=13 \def Ò{\langle}               % Option-[
\catcode211=13 \def Ó{\relax\ifmmode\{\else\ital\fi}      % Option-{
\catcode212=13 \def Ô{\rangle}               % Option-]
\catcode213=13 \def Õ{\}}                        % Option-}
\catcode214=13 \def Ö{\sla}                      % Option-/; Option-V :
\catcode215=13 \def ×{\relax\ifmmode\expandafter\check\else\expandafter\v\fi}
\catcode216=13 \def Ø{\"y}                     % Option-u y
\catcode217=13 \def Ù{\"Y}  		    % Option-u Y
\catcode218=13 \def Ú{\Leftarrow}       % Option-!
\catcode219=13 \def Û{\Leftrightarrow}       % Option-@ ; Option-# :
\catcode220=13 \def Ü{\relax\ifmmode\Rightarrow\else\sect\fi}
\catcode221=13 \def Ý{\sum}                  % Option-$
\catcode222=13 \def Þ{\prod}                 % Option-%
\catcode223=13 \def ß{\widehat}              % Option-^
\catcode224=13 \def à{\pm}                     % Option-&
\catcode225=13 \def á{\nabla}                % Option-(
\catcode226=13 \def â{\quad}                 % Option-)
\catcode227=13 \def ã{\in}               	% Option-W
\catcode228=13 \def ä{\star}      	      % Option-R
\catcode229=13 \def å{\sqrt}                   % Option-M
\catcode230=13 \def æ{\^E}			% Option-i E
\catcode231=13 \def ç{\Upsilon}              % Option-Y
\catcode232=13 \def è{\"E}    	   	 % Option-u E
\catcode233=13 \def é{\`E}               	  % Option-`  E
\catcode234=13 \def ê{\Sigma}                % Option-S
\catcode235=13 \def ë{\Delta}                 % Option-D
\catcode236=13 \def ì{\Phi}                     % Option-F
\catcode237=13 \def í{\`I}        		   % Option-`  I
\catcode238=13 \def î{\iota}        	     % Option-H
\catcode239=13 \def ï{\Psi}                     % Option-J
\catcode240=13 \def ð{\times}                  % Option-K
\catcode241=13 \def ñ{\Lambda}             % Option-L
\catcode242=13 \def ò{\cdots}                % Option-:
\catcode243=13 \def ó{\^U}			% Option-i U
\catcode244=13 \def ô{\`U}    	              % Option-`  U
\catcode245=13 \def õ{\bo}                       % Option-B ; Option-I :
\catcode246=13 \def ö{\relax\ifmmode\expandafter\hat\else\expandafter\^\fi}
\catcode247=13 \def÷{\relax\ifmmode\expandafter\tilde\else\expandafter\~\fi}
\catcode248=13 \def ø{\ll}                         % Option-< ; ^ Option-N
\catcode249=13 \def ù{\gg}                       % Option-> 
\catcode250=13 \def ú{\eta}                      % Option-h 
\catcode251=13 \def û{\kappa}                  % Option-k 
\catcode252=13 \def ü{\half}     		 % Option-Z 
\catcode253=13 \def ý{\Gamma} 		% Option-G 
\catcode254=13 \def þ{\Xi}   			% Option-X ; Option-T : 
\catcode255=13 \def ÿ{\relax\ifmmode{}^{\dagger}{}\else\dag\fi}

      \def\H{{\cal H}}   
    \def\L{{\cal L}}  \def\M{{\cal M}}     
      
\def\S{{\cal S}}      
    
\def\Ä{\varphi}  \def\¿{\varpi}	\def\Ï{\vartheta}

\def\Ç{\textstyle{Ç}}

\begin{document}

\begin{titlepage}
\rightline{September 2015} 
%\rightline{\today} 
\rightline{\tt MIT-CTP-4715} 
\begin{center}
\vskip 2.5cm

{\Large \bf {T-duality Constraints on Higher Derivatives Revisited }}\\

 \vskip 2.0cm
{\large {Olaf Hohm${}^{1,2}$ and Barton Zwiebach${}^1$}}
\vskip 0.5cm
{\it {${}^1$Center for Theoretical Physics}}\\
{\it {Massachusetts Institute of Technology}}\\
{\it {Cambridge, MA 02139, USA}}\\[2ex]
{\it {${}^2$Simons Center for Geometry and Physics}}\\
{\it {Stony Brook University}}\\
{\it {Stony Brook, NY 11794-3636, USA}}\\[2ex]
ohohm@mit.edu, zwiebach@mit.edu\\   

\vskip 2.5cm
{\bf Abstract}

\end{center}

\vskip 0.5cm

\noindent
\begin{narrower}

\baselineskip15pt

We ask to what extent are the higher-derivative 
corrections of string theory constrained by T-duality.  
The seminal early work by Meissner tests T-duality by reduction to one dimension 
using 
a distinguished choice of field variables 
in which the 
bosonic string action takes a Gauss-Bonnet-type form. 
By analyzing all field redefinitions that may or may not be
duality covariant and may or may not be gauge covariant we
extend the procedure to 
test T-duality starting from an action 
expressed in arbitrary field variables. 
We illustrate
the method by showing that it determines
uniquely 
the first-order $\alpha'$ corrections of the bosonic string,
up to terms that vanish in one dimension.   
We also use the method to glean information about
the ${\cal O}(\alpha'^2)$ corrections in the 
double field theory 
with Green-Schwarz deformation.

\end{narrower}

\end{titlepage}

\baselineskip=16pt
\parskip=\medskipamount			% space between paragraphs (TeX)

% \hfill \today

\setcounter{tocdepth}{1}
\tableofcontents

\baselineskip15pt

\sectiono{Introduction}

The T-duality symmetries of string theory have implications for the low-energy
effective field theories obtained after compactification on a torus.  
The full bosonic string theory compactified on a $d$-dimensional torus
has a discrete $O(d,d;\mathbb{Z})$ duality group~\cite{Giveon:1994fu}.
On the other hand, the
low-energy effective field theory for the 
massless degrees of freedom 
has an enhanced $O(d,d; \mathbb{R})$ continuous global symmetry.   This symmetry
was long-recognized in the two-derivative approximation to the effective field theory~\cite{Veneziano:1991ek,Sen:1991zi,9109038}, where the global symmetry transformations 
take a simple form~\cite{Maharana:1992my}.  

More nontrivially, using 
string field theory and the symmetries of
S-matrix elements of massless states it was shown 
in~\cite{Sen:1991zi,9109038} that the
continuous $O(d,d; \mathbb{R})$ global symmetry survives $\alpha'$ corrections
to the effective field theory.  These arguments were recently reviewed and 
elaborated 
to show that the $O(d,d; \mathbb{R})$ 
symmetry in the  
low-energy effective field theory of heterotic strings
is also preserved to all orders in $\alpha'$~\cite{Hohm:2014sxa}.   
Thus the continuous duality symmetry 
is valid for the effective field theory of the full classical string theory.  The arguments prove the existence of the duality symmetry but do not yield the 
associated field transformations.  When the action is written in generally coordinate invariant  
form these dualities 
acquire $\alpha'$ corrections, are rather complicated and are not well understood. 

 The manifest display of global duality symmetries in the presence of $\alpha'$ corrections is a natural subject of study
 in double field theory formulations 
\cite{Siegel:1993th,Siegel:1993bj,Hull:2009mi,Hohm:2010jy,Hohm:2010pp} 
 of the low-energy limits of string theories.
Motivated by the recent progress in encoding 
$\alpha'$ corrections in double field theory  
\cite{Hohm:2013jaa,Hohm:2014eba,Hohm:2014xsa,
Marques:2015vua}  
we revisit here some aspects
of the continuous T-duality symmetry of effective field theories.  

Given an effective field theory for a metric field, a $b$-field, and a dilaton,
one wants to know if this theory has a duality symmetry, by which we mean 
an $O(d,d,\mathbb{R})$ symmetry arising upon dimensional reduction on a torus $T^d$.  If the theory includes higher derivative corrections and is written in generally coordinate invariant notation, the answer is not easily found.  
A test of T-duality to first order in $\alpha'$ was performed by Meissner~\cite{Meissner:1996sa} using the
generally-covariant effective field theory of bosonic 
strings  
and performing a (cosmological) reduction to one dimension. 
In the analysis of~\cite{Meissner:1996sa} it seems necessary to bring this action 
into a particular Gauss-Bonnet-type form by covariant field redefinitions before  
reduction. After reduction one 
aims to rewrite the action 
in terms of a generalized metric and a duality invariant dilaton.  
This final step requires 
further field redefinitions 
that \textit{cannot} originate from covariant redefinitions before reduction.  
The 
test~\cite{Meissner:1996sa} is a necessary condition for T-duality but does not prove it.  An obvious limitation of
this method is that certain linear combinations of terms that are nonzero in arbitrary dimensions
sometimes become zero upon reduction to one dimension.  These combinations
may fail to be T-duality covariant, 
but no constraint arises from the reduction.

A  similar 
analysis of the T-duality constraints on $\alpha'$ corrections was recently   
given by Godazgar and Godazgar~\cite{Godazgar:2013bja}, who consider the reduction on an arbitrary $d$-dimensional torus  but 
truncate to the scalar degrees of freedom, hence giving a necessary but not sufficient condition for T-duality. 
As in the analysis of Meissner, realizing an $O(d,d,\mathbb{R})$ symmetry seems to 
lead to a preferred field basis in the original gravity action. 
It would be useful to have full 
control of the field redefinition freedom.

It is the purpose of this paper to extend the discussion of Meissner to 
make it fully systematic and to deal in all generality with field redefinitions.
Indeed, while the analysis of~\cite{Meissner:1996sa} begins with a `minimal' form of
the ${\cal O} (\alpha')$ effective action,  the method requires the use of covariant
field redefinitions to recast the action in a form where dimensional reduction 
and integration by parts yields terms with no more than 
first-order 
time derivatives.  
This takes a fair amount of work, and the resulting action is significantly more complicated than the original, minimal one. 
Since the existence of T-duality symmetry is independent of field redefinitions, a complete method should work with
the simplest starting point.  
Here we will develop such a method.
Moreover, we also show 
that field redefinitions allow for previously 
unnoticed simplifications of the duality covariant forms of the reduced action.

In the remainder of this introduction we outline our method and results.  
The dimensional reduction for the metric and $b$-field is
based on an ansatz of the form
\be
g_{\mu\nu} \ = \ \begin{pmatrix} -n^2  (t) & 0 \\ 0 & g_{ij} (t) \end{pmatrix}\,,
\qquad    
b_{\mu\nu} \ = \ \begin{pmatrix} 0 & 0 \\ 0 & b_{ij} (t) \end{pmatrix}\,. 
\ee
Here $n(t)$ is the `lapse' function.  One can use the diffeomorphism symmetry
to set $n(t)=1$ but 
one has to remember 
the field equation for $n(t)$.  
This field equation is needed to perform field redefinitions on the reduced action.
After that freedom is taken into account one may set $n(t)=1$, at which point
the action becomes a function of matrices $L, M$, defined  
in terms of the matrices $g_{ij}$ and $b_{ij}$: 
\be
\label{LMdefined}
L \ \equiv \ g^{-1} \dot g \,,  \qquad  M\  \equiv  \ g^{-1}  \dot b  \;. 
\ee
The action will also depend on time derivatives of $L$ and $M$,
as well as on the duality-invariant 
dilaton $\Phi$ and its time 
derivatives.  

In the next step of the procedure one uses the metric and $b$-field equations
of motion to eliminate via field redefinitions any appearance of $\dot L$ and $\dot M$ terms from the action.  Terms with higher time derivatives of $L$ or $M$, if any,
would require integration by parts, until one is able to use the equations of motion. 
We describe identities that allow one to remove  
terms with first derivatives or powers of first derivatives of the dilaton.
Terms with two time derivatives on the dilaton can be eliminated using the dilaton equation to motion and, using integration by parts, so can terms with more than two
time derivatives on the dilaton.    The end result is a simplified reduced 
action that is just a function of traces of powers of $L$ and $M$.  

This result must be set equal to the most general duality invariant one-dimensional
action plus terms that correspond to the `lapse' field redefinition.  Up to equations of motion we demonstrate that the duality covariant action can be written in terms of traces of
powers of first time derivatives of the generalized metric.  
Any term involving  
time derivatives of the dilaton can be redefined away.
  This quickly implies a very simple result: 
the number of parameters in the $2k$-derivative part of the
duality covariant one-dimensional action ($k\geq 2$) is equal to the number $p(k)$ of partitions of $k$.   
The most general lapse field redefinition includes an additional set of parameters.  The test shows T-duality is possible if one can  
adjust  all of those parameters to 
obtain equality with the simplified reduced action. 

We illustrate our method with two examples.  In the first we reconsider the
${\cal O}(\alpha')$ corrections of bosonic string theory.  Up to field redefinitions
these corrections are determined by eight coefficients.  We test T-duality and show
that the correct T-duality covariant action emerges uniquely up to a two-fold ambiguity: there are two linear combinations of terms that are not constrained 
because their reduction to 
one dimension gives zero.\footnote{
For reductions to dimensions $D>1$ these combinations are not zero, and so 
almost surely those linear 
combinations are inconsistent with T-duality.  If that is the case, the ${\cal O}
(\alpha')$ action of the bosonic string is fully determined by continuous
duality.}  
The final $O(d,d)$ covariant four-derivative action at order ${\cal O}(\alpha')$ can be brought to the form
 \be
  S^{(1)} \ = \ \tfrac{1}{16}\int dt  \, e^{-\Phi} \,\Big( {\rm tr}\, \dot{\cal S}^4\, 
  -\, \tfrac{1}{2}\big(  {\rm tr}\, \dot{\cal S}^2\big)^2\Big)\;,   \qquad  {\cal S} \ \equiv \ \eta \H \,,  
 \ee
 where ${\cal H}$ is the generalized metric taking values in $O(d,d)$ 
and $\eta$ is the $O(d,d)$ invariant metric.   
This action is equivalent to that given by Meissner in~\cite{Meissner:1996sa} up to duality covariant 
field redefinitions that eliminate 
all terms with dilaton time derivatives.   

In the second example we consider 
 the `doubled $\alpha'$ geometry' of~\cite{Hohm:2013jaa}.   
 It  has nontrivial
$\alpha'$ corrections and an exact duality symmetry that does not have $\alpha'$
corrections.  In this theory it is the general coordinate transformations that 
receive $\alpha'$ corrections.  Rewritten 
in terms of a conventional metric and $b$-field, however, 
the
duality symmetries will have $\alpha'$ corrections.  Information gathered 
recently~\cite{Hohm:2015mka}  indicates that the ${\cal O}(\alpha')$ corrections cubic in fields are those of a Chern-Simons form based on a torsionless gravitational connection.  We ask here if the simplest form of the action consistent
with this information is T-duality covariant to all orders.  We use
our method to show that this minimal action fails to be T-duality covariant to
${\cal O}(\alpha'^2)$.  
This demonstrates that the 
double field theory of~\cite{Hohm:2013jaa}
must contain additional corrections.

\sectiono{Cosmological reduction and field redefinitions}
In this section we develop 
a method to test if a given action
is consistent with T-duality.  The first step is 
to perform the dimensional reduction 
to one dimension (section 2.1) 
and to bring  that action into 
canonical form (section 2.2).
We discuss 
lapse redefinitions while working in the gauge where
the lapse function is set equal to one.   
We show that,  up to general field redefinitions, 
the action can be written in terms of $L$ and $M$, see (\ref{LMdefined}). In particular, there are no time
derivatives of $L$ or $M$, nor dilaton time derivatives. 
The second step is  to match  the reduced action to a one-dimensional 
duality covariant action, 
whose terms we classify  
up to duality-covariant field redefinitions (section 2.3). 
The matching condition is stated in
equation (\ref{finalduality}).

\subsection{Reduction including 
lapse function and  general field redefinitions}

We begin by performing the (cosmological) reduction to one dimension of 
the standard two-derivative, low-energy action for the bosonic string:  
\be
\label{lavm}
S \ = \   \int d^Dx  \sqrt{-g}\,  e^{-2\phi} \Bigl( R +  4 (\partial \phi)^2 -\tfrac{1}{12} H^2 \Bigr)\;. 
\ee
Here $H_{\mu\nu\rho}  =  3  \partial_{[\mu} b_{\nu\rho]}$ is the field strength
for the $b$-field.  
In the reduction we drop the dependence on all internal coordinates, leaving 
only the dependence on time $t$, 
\be
x^\mu \ = \ (t, x^i) \,, \quad \partial_i \ = \ 0 \,. \ee
For the metric, antisymmetric tensor and (scalar) dilaton we have 
\be
g_{\mu\nu} \ = \ \begin{pmatrix} -n^2  (t) & 0 \\ 0 & g_{ij} (t) \end{pmatrix}\,, \qquad
b_{\mu\nu} \ = \ \begin{pmatrix} 0 & 0 \\ 0 & b_{ij} (t) \end{pmatrix}\,,   \qquad 
\phi \ = \ \phi (t) \,. 
\ee

Before proceeding with the computation of the reduction, it is useful 
to examine the residual diffeomorphisms of the reduction ansatz. 
Since 
we have kept the lapse function $n(t)$, 
we still have time
reparametrization invariance.
This diffeomorphism symmetry $t\rightarrow t-\lambda(t)$ acts as 
 \be
 \begin{split}
  \delta_{\lambda} n \ &= \ \partial_t(\lambda \,n)\;, \\
  \delta_{\lambda} g_{ij} \ &= \ \lambda\, \dot{g}_{ij}\;,  \\
  \delta_{\lambda} b_{ij} \ &= \ \lambda\, \dot{b}_{ij}\;, \\
  \delta_{\lambda} \phi  \ &= \ \lambda\, \dot\phi \;, 
 \end{split}
 \ee  
where we use dots or $\partial_t$ to denote time derivatives.
Note that all fields except for $n(t) $ transform as scalars under time reparameterizations.  The field
$n(t) $ transforms as a density. 
For any field $A$ that transforms as a scalar,  
\be
\delta_\lambda A  \ = \ \lambda \, \partial_t A \,,
\ee 
one can readily verify that the combination $n^{-1} \partial_t$ is
a covariant time derivative and thus $n^{-1} \partial_tA$ is also a scalar:
\be
\delta_\lambda \bigl( n^{-1} \partial_t  A\bigr) \ = \ \lambda \partial_t 
\bigl( n^{-1} \partial_t  A\bigr)\,.
\ee
It is also quickly seen that for any such scalar $A$, the combination $nA$ is
a scalar {\em density}:
\be
\delta_\lambda ( n A)  \ = \ \partial_t ( \lambda n A) \,. 
\ee
It will we useful for us to define a different dilaton by 
\be
\label{lpvm}
e^{-\Phi} \ \equiv \  \sqrt{\det (g_{ij}) } \  e^{-2\phi}  \,.
\ee
Since $g_{ij}$ and $\phi$ transform as scalars, both $e^{-\Phi}$ and
$\Phi$ are scalars:
\be
\delta_\lambda \Phi \ = \ \lambda\  \dot \Phi  \,.   
\ee

\medskip
Let us now begin the calculation of the reduction.  If we reduce (\ref{lavm}) to one dimension, using the definition of the 
dilaton $\Phi$,  
\be
\label{ltsvm}
S \ = \   \ \int dt   \, n  e^{-\Phi}  \Bigl( R +  4 (\partial \phi)^2 -\tfrac{1}{12} H^2 \Bigr)\,.
\ee
In order to compute the various terms in the action, we need the Christoffel symbols, 
whose non-vanishing components are
\be
\Gamma^0_{ij} \ =\  \tfrac{1}{2n^2} \dot g_{ij} \,,  \quad  \Gamma^j_{i0} \ = \
\tfrac{1}{2}  \, g^{jk} \dot g_{ik} \,,  \quad \Gamma^0_{00} \ = \ {\dot n\over n} \,.
\ee
For the lower-index
version  $ \Gamma_{\alpha \mu \nu} \equiv  g_{\alpha\beta} \Gamma^{\beta}_{\mu\nu}$  we get
\be
\Gamma_{0ij}  \ = \ - \Gamma_{ij0} \ = \  -\tfrac{1}{2} \, \dot g_{ij} \,, \quad
\Gamma_{000} \ = \ - n \dot n \,. 
\ee
The non-vanishing components of the Riemann tensor are then found to be 
\be
\label{riemann-pieces}
\begin{split}
  R_{ijkl} \ = \ &  \tfrac{1}{2n^2}\, \dot{g}_{k[i}\,\dot{g}_{j]l}\,,  \\[1.0ex]
   R_{0i0j} \ = \ & -\tfrac{1}{2} n \partial_t ( \tfrac{1}{n} \dot{g}_{ij})  + \tfrac{1}{4}g^{kl} \dot{g}_{ki}\dot{g}_{lj} \,.
 \end{split}
 \ee 
After some calculation using the above, the scalar curvature $R$ is determined to be
\be
R \ = \ \tfrac{1}{n} \partial_t ( g^{ij} \tfrac{1}{n} \partial_t g_{ij})
+ \tfrac{1}{4} \bigl( \tfrac{1}{n} g^{ij} \dot g_{ij} \bigr)^2  
+  \tfrac{1}{4} g^{ij} g^{kl} \tfrac{1}{n} \dot g_{il} \tfrac{1}{n}
\dot g_{kj} \,. 
\ee  
It is manifest that each term here is a scalar. 

It is convenient to define, using matrix
notation,  
\be
L \ \equiv  \ g^{-1} \dot g \,,  \qquad  M\  \equiv  \ g^{-1}  \dot b  \,, 
\ee
so that $L^i{}_j =  g^{ik} \dot g_{kj}$ and  $M^i{}_j = g^{ik} \dot b_{kj}$.  
All our matrices have the row index up and the column index  down.
Note that 
\be
{\cal L} \ \equiv \ n^{-1} L, \quad \hbox{and} \quad 
{\cal M} \ \equiv \ n^{-1} M\,,
\ee
 transform as scalars.  In the gauge $n=1$, ${\cal L}$ becomes $L$ and
 ${\cal M}$ becomes $M$.   Note also the simple identities
  \be\label{ddotsimp}
  g^{-1}\ddot{g} \ = \  \ \dot{L}\, + \, L^2\;, \qquad 
   g^{-1}\ddot{b} \ = \  \ \dot{M}+LM\,. 
 \ee

\medskip
\noindent
Using the above result for the scalar curvature, 
noting that the dilaton relation (\ref{lpvm}) 
gives
\be
\dot \phi \ = \ \tfrac{1}{2} \bigl( \dot \Phi + \tfrac{1}{2} \hbox{tr} L \bigr)  \,, 
\ee
and reducing the kinetic term for the antisymmetric tensor, 
\be
\label{h-sqr}
- \tfrac{1}{12}  H_{\mu\nu\rho}^2 \ = \  - \tfrac{1}{n^2}\, \tfrac{1}{4} \hbox{tr} M^2\,, 
\ee
we find that the two-derivative action in (\ref{ltsvm}) becomes  
\be\label{1D2DerAction}
S \ = \ \int  dt  \, {1\over n} \, e^{-\Phi} \bigl( - \dot \Phi^2  + \tfrac{1}{4} \hbox{tr} (L^2 -M^2) \bigr) \,. 
\ee
The reparameterization invariance is manifest 
because the action equals 
 \be
  S \ = \ \int  dt  \, n \, e^{-\Phi} \bigl( - \big(\,\tfrac{1}{n}\dot \Phi\,\big)^2  + \tfrac{1}{4} \hbox{tr} (\L^2 -\M^2) \bigr) \,,  
\ee
which is written in terms of the covariant time derivatives of dilaton, metric and $b$-field, multiplied 
by the density $n$. 
We find that the metric 
and $b$-field equations of motion take the form
\be
\label{gbfevm}
\begin{split}
\dot{\cal L} \ = \ \,  \partial_t \bigl( n^{-1}  L \bigr)\,  \ = \ & \  n^{-1}
\bigl( \, M^2 + \, \dot \Phi L)  \,, \\[0.5ex]
\dot{\cal M} \ = \ \partial_t \bigl( n^{-1}   M \bigr) \ = \ & \  n^{-1} 
\bigl( ML + \dot \Phi M) \,, 
\end{split}
\ee
while the dilaton equation of motion is
  \be
  \label{dilfevm}
 {d\over dt} \bigl( n^{-1} \dot \Phi\bigr) \ = \ \tfrac{1}{2} \, n^{-1} \, 
 \bigl( \dot \Phi^2 
  + \tfrac{1}{4}  \hbox{tr} (L^2 - M^2) \bigr)  \,.
 \ee 
The equation of motion for the lapse $n$ is quite simple:  it sets the Lagrangian
density equal to zero, which means 
\be
\label{lapsefevm} 
- \dot \Phi^2  + \tfrac{1}{4} \hbox{tr} (L^2 -M^2)  \ = \ 0 \,. 
\ee

We can finally bring the action into manifestly $O(d,d)$ covariant form. 
We first recall that $\eta\H$, where $\eta$ is the invariant metric and $\H$
the generalized metric, takes the form 
\be
\label{etaHdef}
\eta\H \ = \ \begin{pmatrix}  bg^{-1} & g - b g^{-1} b \\[0.7ex]
 g^{-1} & - g^{-1} b 
\end{pmatrix} \;. 
\ee
From this one may verify by a quick calculation that 
\be
\label{kinterm}
\hbox{tr} (\eta \dot \H)^2  \ = \ 2\, \hbox{tr} (M^2-L^2) \,. 
\ee
Comparing with the dimensionally reduced action (\ref{1D2DerAction}), 
one finds that the latter can be written as  
\be
\label{vmsrllct} 
\begin{split}
S \ = \ &  \int dt \, {1\over n} \, e^{-\Phi}  \Bigl(  -\dot \Phi^{\,2} - \tfrac{1}{8}  \hbox{tr} (\eta \dot\H)^2 \,  \Bigr) \,, 
\end{split}
\ee
which is now manifestly $O(d,d)$ invariant.   Both $\Phi$ and $n(t)$ are
inert under $O(d,d)$ transformations.

\subsection{Field redefinitions, lapse gauge fixing and canonical form for the
action}

In order to test if a generally covariant action has T-duality symmetry
we reduce to one dimension.  In this reduction we keep the lapse function
$n(t)$ as a variable.  Since T-duality in two-derivative actions is understood,
the purpose here is to deal with the generally covariant higher derivative
couplings that appear in the effective field theory 
as terms in a power series in $\alpha'$.

Having reduced the full action
the next step is to simplify it in a canonical way by using field redefinitions.
The field redefinitions will be viewed as perturbative in $\alpha'$.  
Our rule will be to use metric, $b$-field and dilaton equations of motion
in order to eliminate {\em all terms} with two or more derivatives of these 
fields.   In order to implement these field redefinitions we can simply
view them as allowed substitutions in the higher-derivative terms.
From (\ref{gbfevm}) and (\ref{dilfevm}) we have the substitutions:
\be
\label{gbfevmbb}
\begin{split}
n^{-1} \dot{\cal L} \ \to  \ & \  n^{-2}
\bigl( \, M^2 + \, \dot \Phi L)  \,, \\[0.5ex]
n^{-1} \dot{\cal M}  \ \to \ & \  n^{-2}  \bigl( ML + \dot \Phi M)\,, \\[0.5ex]
 n^{-1} {d\over dt} \bigl( n^{-1} \dot \Phi\bigr) \ \ \to \ &  \tfrac{1}{2} \, n^{-2} \, 
 \bigl( \dot \Phi^2 
  + \tfrac{1}{4}  \hbox{tr} (L^2 - M^2) \bigr)  \,.
\end{split}
\ee
All objects to the left and right of the arrow are scalars. 
There is one more substitution possible, based on the lapse 
field equation (\ref{lapsefevm}),    
\be
\label{lapsefevm99} 
 \dot \Phi^2 \  \to \  \tfrac{1}{4} \hbox{tr} (L^2 -M^2)  \,. 
\ee
After we use all these substitutions we try to see if the resulting
action has T-duality.  At this point we can use the gauge $n(t)=1$,
and thus the test of T-duality amounts to trying to write the 
resulting action in terms of ${\cal H}$, which encapsulates $g_{ij}(t)$ and
$b_{ij}(t)$, and the dilaton $\Phi$.

In practice we can simplify a bit our work by letting $n=1$  {\em before}
doing the reduction to one dimension and before using the
field equations for the metric, $b$-field and dilaton.  After setting $n=1$
the replacements corresponding to these field equations become
\be
\label{gbfevmbv}
\begin{split}
\dot{L} \ \to  \ & \ 
 \, M^2 + \, \dot \Phi L  \,, \\[0.5ex]
 \dot{ M}  \ \to \ & \   ML + \dot \Phi M \,, \\[0.5ex]
\ddot \Phi \ \ \to \ & \  \tfrac{1}{2} 
 \bigl( \dot \Phi^2 
  + \tfrac{1}{4}  \hbox{tr} (L^2 - M^2) \bigr) \ \to \ 
   \tfrac{1}{4}  \hbox{tr} (L^2 - M^2)  \,,  
\end{split}
\ee
where we used the lapse equation in the last replacement.  
Since in this way we lose the
$n$ field equation, we must recall that we have the
ability to do the lapse-related field redefinition (\ref{lapsefevm99}).  
Setting $n=1$ before reduction and use of  the equations of motion
gives the same result as setting $n=1$ after reduction and use
of equations of motion.  This is because all terms in the action are
time-reparameterization scalars and for them {\em all} 
appearances of $n$ are in the form $n^{-1} \partial_t$.  Setting $n=1$
then just leaves the time derivatives that would have been obtained otherwise.

Having removed terms with two derivatives on the fields another
important set of identities allows 
us to remove terms with  powers of
$\dot \Phi$.  Assume $X= X (L,M)$ is a function of $L$ and $M$ only and 
is of degree $k$:  
$X (\lambda L, \lambda M) = \lambda^{k_X} X(L,M)$. 
 We  can then manipulate a term of the form
$\dot \Phi  \, X$ in the Lagrangian as follows:
\be
\int dt e^{-\Phi} \dot \Phi \, X \ = \ -\int dt \, \partial_t (e^{-\Phi}) \, X 
 \  = \ \int dt e^{-\Phi}  \, \dot X\,.
\ee
Given that $X$ is of degree $k_X$, if we take the time derivative and
use the substitutions in (\ref{gbfevmbv}) to eliminate the $\dot L$ and
$\dot M$ terms we will find
\be
\dot X  \ = \ k_X \, \dot \Phi \, X  + X'\,,
\ee
where prime denotes the action of taking a time derivative
 and letting $\dot L \to M^2$ and $\dot M \to ML$. 
 As a result we have 
 \be
\int dt\, e^{-\Phi} \dot \Phi \, X 
 \  = \ \int dt\, e^{-\Phi}  
 \, ( k_X \, \dot \Phi \, X  + X') \,,
\ee
which means that as terms in the Lagrangian we eliminate the $\dot\Phi X$ 
term via the equivalence
\be
\label{dil-dot-rep-vm}
\dot \Phi \, X \ \simeq   \   {1\over (1-k_X)}  \,  X' \,.
\ee
Here and in the following we denote by $\simeq$ equalities that hold up to 
equations of motion and integrations by parts. 
The above identity  fails for $k_X =1$, but  this case will not be relevant,
as it would correspond to a term with two derivatives and we are interested
in higher-derivative terms. 
As examples of the use of this identity consider two forms of $X$, both of degree
three:   
  \be\label{totderidentities}
   \begin{split}
    \dot{\Phi}\,{\rm tr}(L^3) \ &\simeq \ -\tfrac{3}{2}\, {\rm tr}(M^2L^2)\;, \\
    \dot{\Phi}\,{\rm tr}(M^2L) \ &\simeq \ -\tfrac{1}{2}{\rm tr}(MLML+M^2L^2+M^4)\;. \\
     \end{split}
  \ee  
By a similar analysis, this time using the dilaton replacement
 (last line in (\ref{gbfevmbv})), one can show that   
 \be\label{dilatonIDentities-99-vm}
\begin{split}
\dot \Phi^2 \, X \ \simeq  \  -{1\over 2k_X-1} \bigl( \, \tfrac{1}{4} 
\hbox{tr} (L^2 -M^2) X  \, - \, \tfrac{2}{k_X}  \,  X'' \bigr) \,.
 \end{split}
 \ee
Here $X'' = (X')'$, using the definition of prime given above. Since 
$k_X =0$ is not of interest and $k_X$ is an integer, the above formula
always gives a well defined equivalence. 
As  examples we record 
  \be\label{totderidentities-vm}
   \begin{split}
    \dot{\Phi}^{\,2}\,{\rm tr}(L^2) \ & \simeq  \ -\tfrac{1}{12}\,{\rm tr}(L^2-M^2)\,{\rm tr}(L^2)
    +\tfrac{2}{3} 
    {\rm tr}(MLML+M^2L^2+M^4)\,, \\ 
       \dot{\Phi}^{\,2}\,{\rm tr}(L^2 - M^2) \ & \simeq  \ -\tfrac{1}{12}\,\bigl( {\rm tr}(L^2-M^2)\bigr)^2 \;. 
 \end{split}
  \ee  
For arbitrary  
powers of $\dot\Phi$ we can use the following relation,
derived by exactly the same methods:
\be
\label{recursive-vm}
\tfrac{1}{2} \bigl( 3- p - 2k_X \bigr) 
\, \dot \Phi^p \, X \ \simeq \ 
\tfrac{p-1}{8} \, \hbox{tr} (L^2-M^2)  \, \dot\Phi^{p-2} \, X  \ + \dot \Phi^{p-1} X' 
 \,. 
\ee
This relates a term with 
$\dot \Phi^p$ to terms with $\dot \Phi^{p-1}$
and $\dot \Phi^{p-2}$ and can be used recursively. 
Note that  (\ref{dil-dot-rep-vm}) follows  from for $p=1$, and
(\ref{dilatonIDentities-99-vm}) follows
for $p=2$, after using the $p=1$
result. 
This equation shows
 that we can always eliminate the $\dot\Phi$ dependence
of terms using field equations.   The prefactor on the left-hand side indicates
that for $p=1, k_X=1$  or for $p=3, k_X =0$ the relation fails to help
eliminate $\dot \Phi^pX$. 
But those cases correspond to terms with two and three derivatives, respectively,
and are of no interest to us.  
 The upshot of this analysis is that, 
 by the use of field redefinitions,
 the dimensionally
reduced action  can be written as a function of 
$L$ and $M$ with no extra time derivatives and without any dilaton time derivatives.

As we mentioned while introducing (\ref{lapsefevm99}),  
we simplify the reduced action using  
the lapse equation of motion    
to replace, for arbitrary $Y$,   
\be\label{lapseREDEf}
\dot \Phi^2 \, Y \ \simeq \  \tfrac{1}{4} \hbox{tr} (L^2 -M^2)  \, Y \,.
\ee
We could use (\ref{dilatonIDentities-99-vm}) instead, but this is
more complicated and not needed at this stage.  
In order to explain this point, let us consider the 
general ansatz that needs to be solvable in order for duality invariance 
to be possible.  
Schematically, the matching equation reads  
 \be\label{finalduality}
  (\text{reduction to $D=1$)} \,\,\, \simeq  
  \,\,\, \text{(general 1D duality-invariant action)}  
  \,+\,\Bigl( \dot \Phi^2 - \tfrac{1}{4} \hbox{tr} ( L^2 -M^2) \Bigr)  X \,, 
 \ee  
where the last term accounts for lapse redefinitions and
 $X$ is an arbitrary function of $L, M,$ and $\dot\Phi$. 
 This form makes it manifest that the use of (\ref{lapseREDEf}) is
 legal in the simplification of the left-hand side of the matching equation. 
 Indeed,  a term  $\dot \Phi^2 \, Y$ on the left-hand side can be trivially 
 rewritten as 
 \be
 \dot \Phi^2 \, Y \ = \  \tfrac{1}{4} \hbox{tr} (L^2 -M^2)  \, Y  
 + (\dot \Phi^2- \tfrac{1}{4} \hbox{tr} (L^2 -M^2)) Y \,, 
 \ee
and we can ignore the second term as long as $X$ is general.
If the reduction to 1D with all its simplifications has been carried out,
and given that the general 1D duality-invariant action has no dilaton
time derivatives (section 2.3), the only dilaton time derivatives in (\ref{finalduality})
are in the second term of the right-hand side.   At this point 
the equivalence (\ref{dilatonIDentities-99-vm}) and, more generally, (\ref{recursive-vm}), both valid up to field redefinitions,  
  is needed to eliminate such dependence.  
This is why we use the symbol $\simeq$ in (\ref{finalduality})

To illustrate the simplification procedure for the reduction of the action to $D=1$,
consider the reduction of Riemann-squared.
Using the non-vanishing components of the Riemann
tensor one quickly finds 
\be
R_{\mu\nu\rho\sigma}^2 \ = \ R^{ijkl} R_{ijkl} + 4 (g^{00})^2  R_0{}^j{}_{0i}
R_0{}^i{}_{0j}\,.
\ee
It follows immediately from the first equation in (\ref{riemann-pieces}) that 
\be
 R^{ijkl} R_{ijkl} \ = \ \tfrac{1}{8} ( \hbox{tr}\, {\cal L}^2)^2  - \tfrac{1}{8} \, \hbox{tr} \, {\cal L}^4 \,. 
\ee
Moreover,  a few lines of calculation using the second equation in 
(\ref{riemann-pieces}) shows that
\be
R_0{}^i{}_{0j} \ = \ - \tfrac{1}{2} \, n^2  \bigl( n^{-1} \dot {\cal L} + \tfrac{1}{2} {\cal L}^2 \bigr)^i{}_j \,. 
\ee
All in all Riemann-squared gives
\be
\begin{split}
R_{\mu\nu\rho\sigma}^2 \ = \  & \ \tfrac{1}{8} ( \hbox{tr}\, {\cal L}^2)^2  - \tfrac{1}{8} \, \hbox{tr} \, {\cal L}^4 + \, \hbox{tr} \bigl(   n^{-1} \dot {\cal L} + \tfrac{1}{2} {\cal L}^2 \bigr)^2 \\[1.0ex]
\ = \ & \  \hbox{tr} \bigl( \, (n^{-1}  \dot {\cal L})^2  +  n^{-1} \dot {\cal L} {\cal L}^2 
 + \tfrac{1}{8} \, {\cal L}^4 \bigr)  + \tfrac{1}{8} ( \hbox{tr}\, {\cal L}^2)^2 \,.
\end{split}
\ee
This is manifestly a scalar and, as expected, the time derivatives always appear
accompanied by a factor of $n^{-1}$.   Using the equation of motion for $g_{ij}$
and setting $n=1$ 
afterwards  
gives manifestly the same result as setting $n=1$ first
and then using the simpler equations of motion.  Setting first $n=1$
we have 
\be
\begin{split}
R_{\mu\nu\rho\sigma}^2 \ = \ & \  \hbox{tr} \bigl( \, \dot { L}^2  + \dot {L} {L}^2 
 + \tfrac{1}{8} \, {L}^4 \bigr)  + \tfrac{1}{8} ( \hbox{tr}\, { L}^2)^2 \,.
\end{split}
\ee
Using now the replacement associated to the equation of motion (\ref{gbfevmbv}) 
we get
 \be
 \begin{split}
 R_{\mu\nu\rho\sigma}R^{\mu\nu\rho\sigma} \ \simeq  \  \ 
 &  \hbox{tr} \bigl(  \tfrac{1}{8} L^4
 + M^2 L^2 + M^4 \bigr)  +  \tfrac{1}{8}  \bigl( \hbox{tr} (L^2) 
 \bigr)^2 \\[0.8ex]
 & \ +   \dot \Phi \, \hbox{tr} (L^3 + 2M^2L)  + 
 \dot \Phi^2 \, \hbox{tr} (L^2)\,.
 \end{split} 
 \ee 
Employing next (\ref{totderidentities}) and the lapse equation replacement we get
 \be
 \begin{split}
R_{\mu\nu\rho\sigma}R^{\mu\nu\rho\sigma}  \ \simeq \  \ 
 &  \hbox{tr} \bigl(  \tfrac{1}{8} L^4
 + M^2 L^2 + M^4 \bigr)  +  \tfrac{1}{8}  \bigl( \hbox{tr} (L^2) 
 \bigr)^2 \\[0.5ex]
 & \ +  \, \hbox{tr} (-MLML -\tfrac{5}{2} M^2 L^2  -M^4)  + 
 \tfrac{1}{4} \hbox{tr} (L^2-M^2) \, \hbox{tr} (L^2) \,. 
 \end{split}
 \ee
Simplifying, we finally get:
 \be
 \label{riemann-sqr-simp}
R_{\mu\nu\rho\sigma}R^{\mu\nu\rho\sigma}  \ \simeq \  \ 
   \hbox{tr} \bigl(  \, \tfrac{1}{8} L^4\, 
 -MLML\, -\tfrac{3}{2} M^2 L^2  \bigr)  +  \tfrac{3}{8}  \bigl( \hbox{tr}\, L^2 
 \bigr)^2\,  - 
 \tfrac{1}{4} \,\hbox{tr} M^2 \, \hbox{tr} L^2 \,, 
 \ee
where we recall that  $\simeq$ means that  the left-hand side and right-hand side
are equal up to field redefinitions and integrations by parts.  
This expression will be needed in the later analysis.

%\medskip
Since we can do the reduction to 
one dimension without using the
lapse function we collect 
a few formulae.
The nonvanishing Christoffel symbols, $b$-field field strengths and 
curvatures are
\be
\Gamma^0_{ij} = \tfrac{1}{2} \dot g_{ij} \,,  \quad  \Gamma^j_{i0} \ = \
\tfrac{1}{2}  \, g^{jk} \dot g_{ik} \,,  \quad \hbox{or} \quad
\Gamma_{0ij}  \ = \ - \Gamma_{ij0} \ = \  -\tfrac{1}{2} \, \dot g_{ij} \,. 
\ee
\be
 H_0{}^i{}_j \ = \ (M)^i{}_j \,,   \quad     - \tfrac{1}{12}  H_{\mu\nu\rho}^2  \ = \ -\tfrac{1}{4} \hbox{tr} M^2 \,. 
\ee
\be
  R_{ijkl} \ = \   \tfrac{1}{2}\,\dot{g}_{k[i}\,\dot{g}_{j]l}\,, \qquad 
   R_0{}^i{}_{0j} \ = \  -  \tfrac{1}{2} \bigl( \dot L   + \tfrac{1}{2} L^2 \bigr)^i{}_j \,, 
 \ee 
 \be
 R_{00} \ =  \, -\tfrac{1}{2} \, \hbox{tr} (\dot L)  - \, \tfrac{1}{4} \, \hbox{tr} (L^2) 
 \,, \quad 
 R^i{}_j \ =  \, \tfrac{1}{2} (\dot L  + \tfrac{1}{2}  \, L \,\hbox{tr} L)^i{}_j \,, \quad 
 R \ = \,   \hbox{tr} (\dot L)  +  \tfrac{1}{4} ( \hbox{tr} L )^2  + \tfrac{1}{4} 
  \,\hbox{tr} (L^2) \,.  
 \ee

To check the consistency of the reduction and our formulae
we have examined in detail the covariant field equations for
the metric, $b$-field, and dilaton.   
Their  reduction give the equations of
motion displayed above, with the lapse equation arising
from the $g_{00}$ equation.
We have also checked that we are not missing nontrivial field
equations by setting $g_{0i}=0$ and $b_{0i}=0$ 
in the reduction ansatz.

\subsection{$O(d,d)$ covariant field redefinitions}

In this subsection we consider the duality covariant, one-dimensional,
two-derivative 
action for ${\cal H}$ and $\Phi$ and then examine what are the
possible duality covariant 
$\alpha'$ corrections, up to field redefinitions.   The 
action, setting $n=1$ in (\ref{vmsrllct}), is 
\be
\label{vmsrllctij} 
\begin{split}
S \ = \ &  \int dt \,\, e^{-\Phi}  \bigl(  -\dot \Phi^{\,2} - \tfrac{1}{8}  \hbox{tr}\, \dot {\cal S}^2 \,  \bigr) \,, 
\quad {\cal S} \ \equiv \ \eta  \H  \, .  
\end{split}
\ee
For brevity we have introduced $\S = \eta \H$, 
which satisfies $\S^2=1$. 
The equations of motion for 
$\S$ and $\Phi$ are then 
\be
\begin{split}
\ddot{\S}+\S\dot{\S}^2-  \,\dot{\Phi}\,\dot{\S}\ = \ &  \ 0 \,, \\[0.5ex] 
\ -2\ddot{\Phi}+\dot{\Phi}^{\,2}-\tfrac{1}{8}{\rm tr}\,\dot{\S}^2 \ = \ & \  0\,.
\end{split}
\ee 
Thus, using $O(d,d)$ covariant field redefinitions we can always replace 
 \be
 \label{cov-replacements}
 \begin{split}
  \ddot{\S}\,\rightarrow & \ \, -\S\dot{\S}^2 + \,\dot{\Phi}\,\dot{\S}\,, \\[0.5ex]
 \ddot{\Phi}\,\rightarrow & \ \,\  \tfrac{1}{2}\, \dot{\Phi}^2 - \tfrac{1}{16}  \hbox{tr}\, \dot \S^2 \;. 
\end{split}
 \ee

\medskip
We will next apply this freedom of duality covariant field redefinitions 
in the one-dimensional ${\cal O}(\alpha')$ action found by Meissner. 
It is given by $\int dt  e^{-\Phi} {\cal L}$ with Lagrangian 
\cite{Meissner:1996sa}
 \be
  {\cal L} \ = \  \tfrac{1}{16}\, {\rm tr}\, \dot \S^4-\tfrac{1}{64}\big({\rm tr}\, \dot \S^2\big)^2
  -\tfrac{1}{4}\, \dot{\Phi}^{\,2}\, {\rm tr}\,\dot \S^2\, -\tfrac{1}{3}\,\dot{\Phi}^{\,4}\, .
 \ee 
We will see that all dilaton terms can be removed by $O(d,d)$ invariant field redefinitions. 
We recall relation (\ref{recursive-vm}), which in $O(d,d)$
covariant language reads
\be
\label{recursive-vm-pb}
\tfrac{1}{2} \bigl( 3- p - 2k_X \bigr) 
\, \dot \Phi^p \, X \ \simeq \ 
-\tfrac{p-1}{16} \, \hbox{tr}\, \dot \S^2 \, \dot\Phi^{p-2} \, X  \ + \dot \Phi^{p-1} X' \,.
\ee
For $X$ equal to a constant we get
\be
\label{pure-dil-vm}
\dot \Phi^k \ \simeq \  \tfrac{1}{8} \, \bigl(\tfrac{ k-1}{k-3}\bigr)  \, \dot \Phi^{k-2} \, \hbox{tr}\, \dot \S^2 \,.  
\ee
This allows us to trade the $\dot \Phi^4 $ term in the above Lagrangian for
a $\dot \Phi^2 \, \hbox{tr} \, \dot \S^2$ 
term.\footnote{Note, however, that we could not eliminate a $\dot\Phi^3$ term -- but this is 
irrelevant 
because such term has three derivatives and is of no interest to us.}  
Note now that the second relation in (\ref{totderidentities-vm}) implies that
\be
\dot \Phi^2 \, \hbox{tr} \, \dot \S^2  \ \simeq \   \tfrac{1}{24}  \,  \bigl( \hbox{tr} \, 
\dot \S^2 \bigr)^2 \,. 
\ee
Using these relations one quickly shows that the above
Lagrangian (viewed as an addition to the two-derivative theory) 
is field-redefinition equivalent to the simpler
 \be\label{MeissnerImproved}
  {\cal L} \ \simeq  \ \tfrac{1}{16}\,
  {\rm tr}\, \dot \S^4\, -\, \tfrac{1}{32}\, \big({\rm tr}\, \dot \S^2\big)^2\,. 
 \ee

\subsubsection*{Classification of $O(d,d)$ invariants}

We will classify all possible $O(d,d)$ invariant terms that can be
added to the two-derivative action.  These will be written in terms
of $\S = \eta \H$, its derivatives, the dilaton $\Phi$ and its derivatives.
Terms that differ by $O(d,d)$ covariant field redefinitions will be considered
equivalent.  
The result is simple:  terms are constructed by taking traces, or products
of traces, of even powers of $\dot \S$.  In particular, terms involving
the dilaton time derivatives do not appear.
 
To begin note that $\S$ has zero trace (see (\ref{etaHdef})),
and therefore so do all of its derivatives, 
\be
\hbox{tr} (\S) \ = \ 
\hbox{tr} (\dot \S) \ = \ \hbox{tr} (\ddot \S) \ = \ 0 \,.
\ee
Moreover, since $ \S\S = 1$ we immediately learn that $\S$ and $\dot \S$ 
anticommute:
\be\label{anticommute}
\S \dot \S + \dot \S \S \ = \ 0\,.
\ee
We first show that traces of odd powers of $\dot \S$ vanish. 
For this purpose we take a second derivative of the above equation to get
\be\label{ddotidentity}
2 \dot \S \dot \S  +  \ddot \S \S  + \S \ddot \S  \ = \ 0 \,. 
\ee
Multiplying from the left by  $(\dot \S)^{2k+1}$, with $k$ a non-negative
integer,  we find
\be\label{ddotidentity00}
2 (\dot \S)^{2k+3}  +  (\dot \S)^{2k+1}\ddot \S \S  + (\dot \S)^{2k+1}\S \ddot \S  \ = \ 0 \,.
\ee
Taking traces and using cyclicity, we have
\be\label{ddotidentity998}
2 \,\hbox{tr} (\dot \S^{2k+3})  + \hbox{tr} \bigl( \S\,\dot \S^{2k+1}\ddot \S   
+  \dot \S^{2k+1}\, \S \ddot \S\bigr)   \ = \ 0 \,.
\ee
Noting now with (\ref{anticommute}) 
that $\S \dot \S^{2k+1} = - \dot \S^{2k+1}\S$, we learn
that 
\be
\hbox{tr} (\dot \S^{2k+3} ) \ = \ 0 \,.
\ee
Since $\hbox{tr} (\dot \S)=0$, we have now proven, as we claimed, that 
\be
\label{oddpowertrace} 
\hbox{tr} (\dot \S^{2k+1} ) \ = \ 0 \,,  \hbox{   for} \quad k=0,1, \ldots \,. 
\ee

There is no need to consider the use of $\ddot \S$ or higher time 
derivatives of $\S$.  Using the $\S$ field equation we can
implement the replacements in (\ref{cov-replacements}) to trade a double
derivative of $\S$ for terms with $\S, \dot \S$ and $\dot \Phi$.  
We
see that we must consider terms that also involve 
the undifferentiated $\S$. 
There is nothing that can be done with just $\S$, as it has zero trace, and
once squared it equals the identity matrix.  The question is if we can
build some new duality invariant using $\S$ and $\dot \S$.  The answer
is no, as we show next.

If we have a trace of a string of products of $\S$'s and $\dot \S$'s, using the 
anticommutativity 
of  $\S$ and $\dot \S$ the term can be arranged so that all the $\S$'s are 
near each other and thus, since $\S^2=1$, the only possible terms are
of the form
\be \hbox{tr} ( \S \dot \S^{k}) \,,
\ee 
for $k$ a non-negative integer.   
It is straightforward to see that they vanish for all $k$,  including $k=1$,   
 \be
  \hbox{tr} ( \S \dot \S^{k}) \ = \ \hbox{tr} (\S \dot\S \dot \S^{k-1}) \ = \ -\,\hbox{tr} (\dot \S \S \dot \S^{k-1})
  \ = \ -\, \hbox{tr} ( \S \dot \S^{k}) \ = \ 0\;. 
 \ee
Here we used  
(\ref{anticommute}) in the second equality 
and cyclicity of the trace in the third equality.

The last issue we have to discuss is terms with time derivatives of
the dilaton.  Only first derivatives are relevant, since terms with two
or more time derivatives of the dilaton can be reduced by using the
dilaton field equation.   But we have already seen that (\ref{recursive-vm-pb})
allows us to get rid of such terms.  Therefore there are no dilaton time derivatives
in the one-dimensional duality invariant action, up to field redefinitions.

Our analysis implies that for four derivatives the 
most general duality covariant terms that, up to field redefinitions,
can be added to the action are
\be
\label{fourder}
{\cal O} (\alpha'): \quad  a_1  \hbox{tr} \dot \S^4\,    + a_2  
   \bigl( \hbox{tr} \dot \S^2\bigr)^2 \,.
\ee
For six derivative terms we have
\be
\label{six-der} 
{\cal O} ((\alpha')^2): \quad  
\ c_1\,{\rm tr}\,\dot \S^6\ +c_2\,{\rm tr}\,\dot \S^4\,{\rm tr} \, \dot \S^2
+c_3 \big({\rm tr} \, \dot \S^2\big)^3\,. 
\ee
The terms that arise at order $(\alpha')^{k-1}$ have $2k$ derivatives and look like
\be
\label{six-der} 
{\cal O} ((\alpha')^{k-1}) \, : \quad 
\ c_1\,{\rm tr}\,\dot \S^{2k} \ +c_2\,{\rm tr}\,\dot \S^{2k-2}\,{\rm tr} \, \dot \S^2
+ c_3 \, {\rm tr} \, \dot \S^{2k-4} (\hbox{tr} \,\dot \S^2)^2  
+c_4 \, {\rm tr} \, \dot \S^{2k-4} \, \hbox{tr} \, \dot \S^4 + \ldots 
\ee
Letting  $z \equiv \dot\S^2$ and assuming that each factor in
a product has its trace taken, the above expression is
\be
c_1 \, z^{k}  + c_2  z^{k-1} z  + c_3 \, z^{k-2}  z\, z   +  c_4 \, z^{k-2}  z^2 
+ \ldots\,,
\ee
making it clear that each summand can be associated with a partition of the integer $k$.
Thus the number of independent coefficients in the ${\cal O} ((\alpha')^{k-1})$
action is $p(k)$,  the number of partitions of $k$.  This is the number of coefficients
in the part of the action with $2k$ derivatives.

For reference we 
collect the first few independent invariants in terms of $L$ and $M$, 
\be\label{OddTraces}
\begin{split}
\hbox{tr} (\dot \S^2) \ &= \ 2 \, \hbox{tr} \,\bigl(\,    -  L^2 + M^2 )  \;, \\
\hbox{tr} (\dot \S^4)\ &= \ 2 \, \hbox{tr} \,\bigl( \,   L^4 + 2 M LML
- 4 M^2 L^2 + M^4  )\;,  \\
\hbox{tr} (\dot \S^6)  \ &= \ 2 \, \hbox{tr} \,\bigl(\,   - L^6   - 6 ML^3 ML \, + 3ML^2 ML^2   + 6 M^2 L^4  \\
 & \hskip+40pt - 3 M^2 LM^2 L   + 6 M^3 LML - 6 M^4 L^2+  M^6 \bigr) \;. 
\end{split}
\ee

\sectiono{Bosonic string theory at ${\cal O}(\alpha')$ revisited}

As an application  and illustration 
of the general procedure developed above, we 
consider  
the bosonic string effective action including ${\cal O}(\alpha')$ corrections 
and investigate to what extent it is constrained by duality invariance in 
the reduction to one dimension. 
We start with the most general four-derivative action, up
 to field redefinitions, according to Tseytlin and Metsaev.  This action
 contains
 eight terms and thus eight coefficients $\gamma_i,  \, i = 1, \ldots , 8$,  
 and takes the form~\cite{Metsaev:1987zx}:
 \be
 S(\gamma) \ = \ \int  d^D x \sqrt{-g} \,  e^{-2\phi} \, {\cal L}(\gamma) \;, 
 \ee
 where
 \be\label{Lofgamma}
 \begin{split}
{\cal L} (\gamma) \ =  \ & \ 
 \gamma_1 \, R_{\mu\nu\rho\sigma}^2  
 \  +\ \gamma_2 \,  HHR 
 \  +\ \gamma_3 \, H^4  
 \ +\ \gamma_4 \,  (H^2_{\mu\nu} )^2  \\[1.0ex]
&  \ +\  \gamma_5 \,   (H^2)^2 
 \ +\ \gamma_6 \,  H^2_{\mu\nu} \, \partial^\mu\phi  \partial^\nu\phi 
  \ +\ \gamma_7\,  H^2 (\partial \phi)^2  
  \ +\ \gamma_8 \, (\partial \phi)^4    \,.
\end{split}
\ee
In here, the various terms are defined as follows: 
   \be
 \begin{split}
 HHR \ \equiv \ & \    H^{\mu\nu\lambda}H^{\rho\sigma}{}_{\lambda} R_{\mu\nu\rho\sigma} \;,  \\[0.3ex]
 H^4 \ \equiv \ & \ H_{\mu\nu\rho} 
(HHH)^{\mu\nu\rho} \ \equiv \  H_{\mu\nu\rho} \
H^{\mu}{}_\alpha{}^\beta H^{\nu}{}_\beta{}^\gamma
H^{\rho}{}_\gamma{}^\alpha \;, \\[0.8ex]
\quad H^2_{\mu\nu} \ \equiv \ &  
H_{\mu}{}^{\alpha\beta} H_{\nu}{}_{\alpha\beta} \,,\\[0.8ex]
 (H^2_{\mu\nu})^2 \ \equiv \ &  \ H^2_{\mu\nu} H^{2\, \mu\nu}\;,  \\[0.2ex]
 H^2 \ \equiv \  & \  H_{\mu\nu\rho}H^{\mu\nu\rho} \;. 
  \end{split}
 \ee 
Given the 
known result that the ${\cal O}(\alpha')$ action of 
bosonic string theory,
 up to field redefinitions, is   \cite{Metsaev:1987zx}
   \be
\int  d^D x \sqrt{-g} \,  e^{-2\phi}\, \Bigl( R_{\mu\nu\rho\sigma}^2  - \tfrac{1}{2}  HHR  + \tfrac{1}{24} \, H^4  - \tfrac{1}{8} (H^2_{\mu\nu})^2 \, \Bigr) \,,
\ee
we know that 
 \be
 \gamma_1 = 1 , \ \ \gamma_2 = -\tfrac{1}{2} \,, \ \ 
 \gamma_3 = \tfrac{1}{24} \,, \ \  \gamma_4 = -\tfrac{1}{8} \,,   \ \ 
 \gamma_5 = \gamma_6 = \gamma_7 = \gamma_8 =0 \,,
 \ee
 is a duality invariant solution.  We now want to see if this
 is the answer selected by the condition of duality invariance.  

The strategy is to reduce the Lagrangian ${\cal L}(\gamma)$ 
down to one dimension
using the field equations to eliminate $\dot L, \dot M, \ddot \Phi$ and
$\dot \Phi^2$ terms.   The result is equated to the general
duality covariant terms plus the most general lapse 
redefinition, as explained in (\ref{finalduality}).  
The lapse redefinition takes the form
\be
\Bigl( \dot \Phi^2 - \tfrac{1}{4} \hbox{tr} ( L^2 -M^2) \Bigr)  
 \bigl( b_1 \hbox{tr} (L^2) + b_2 \hbox{tr} (M^2) 
   + b_3 (\hbox{tr}\,  L)^2   + 
   b_4 \dot \Phi^2 + b_5 \dot\Phi (\hbox{tr} L) \bigr) \;,  
\ee
where the terms in parenthesis are the most general terms we can write
with two time derivatives, realizing that  $\hbox{tr} (M) = \hbox{tr} (ML) =0$.
The term with coefficient $b_5$ gives,   
up to equations of motion, 
precisely minus the contribution of the term with 
coefficient $b_2$.  This is so because, up to equations of motion,  
\be
e^{-\Phi} \bigl( \dot \Phi^2 - \tfrac{1}{4} \hbox{tr} ( L^2 -M^2) \bigr)  
 \hbox{tr} (M^2 + \dot \Phi L)  \ \simeq \ 
 e^{-\Phi} \bigl( \dot \Phi^2 - \tfrac{1}{4} \hbox{tr} ( L^2 -M^2) \bigr)  
 \, \hbox{tr} \, \dot L \,, 
  \ee
and the term in the right hand side is a total 
total derivative, as can be checked integrating by parts the time derivative
acting on $L$ and 
using the dilaton, metric, and $b$-field equations of motion.  
Therefore we can set $b_5$ equal to zero.   
Using (\ref{fourder}) and (\ref{finalduality}) the dimensionally reduced Lagrangian 
should then  be writable as 
\be
\label{vm-vm!}
\begin{split}
{\cal L}(\gamma) \bigl|_{1d}  \ \simeq    
 \ &    \   a_1  \hbox{tr} (\eta \dot \H)^4   + a_2  
   \bigl( \hbox{tr} (\eta \dot \H)^2\bigr)^2  \\[1.0ex]
&     +  \Bigl( \dot \Phi^2 - \tfrac{1}{4} \hbox{tr} ( L^2 -M^2) \Bigr)  
   \bigl( b_1 \hbox{tr} (L^2) + b_2 \hbox{tr} (M^2) 
   + b_3 (\hbox{tr}\,  L)^2   + b_4 \dot \Phi^2  \bigr) \;,   
\end{split}
\ee 
which will constrain the $\gamma$ coefficients. 
This is our key equation.  The right-hand side can be evaluated 
with (\ref{OddTraces}) 
and (\ref{dilatonIDentities-99-vm}) 
to  give
\be
\label{rhs-vm}
\begin{split}
{\rm rhs} \ \simeq   
 \  &  \    \hbox{tr} \,\bigl( \ 2 a_1 \, L^4 + 4a_1\, M LML
- 8a_1\, M^2 L^2 +2a_1\, M^4  )  \\[1.0ex]
&     + 4a_2 \, (\hbox{tr} L^2)^2  + 4a_2 \, (\hbox{tr} M^2)^2  
- 8a_2 \, (\hbox{tr} M^2)(\hbox{tr} L^2)  \\[1.0ex]
&  + b_1 \Bigl( -\tfrac{1}{3} \hbox{tr} (L^2-M^2)\,  \hbox{tr} L^2 \, + \tfrac{2}{3}
\hbox{tr} \bigl( MLML + M^2 L^2 + M^4  \bigr) \Bigr) \\[1.0ex]
&  +  b_2 \Bigl( -\tfrac{1}{3} \hbox{tr} (L^2-M^2)\,  \hbox{tr} M^2 \, + \tfrac{2}{3}
\hbox{tr} \bigl( MLML + M^2 L^2 + M^4  \bigr) \Bigr) \\[1.0ex]
&  + b_3 \Bigl( -\tfrac{1}{3} \hbox{tr} (L^2-M^2)\,  (\hbox{tr}\, L)^2 \, + \tfrac{2}{3}
\, ( \hbox{tr} \, M^2)^2  + \, \tfrac{4}{3} \hbox{tr} (M^2L)  \, \hbox{tr} \, L 
 \Bigr) \\[1.0ex]
& +b_4\, \tfrac{1}{12} \,   (\hbox{tr} (L^2 - M^2))^2  \;.  
\end{split}
\ee 

 Now we must evaluate the left-hand side 
 by computing the cosmological reduction of (\ref{Lofgamma}).   The Riemann-squared term was given in (\ref{riemann-sqr-simp}).
  The HHR invariant yields 
 \be
 \begin{split}
 HHR \ \ &= \ 
  {\rm tr}\big(M^2L^2+\tfrac{1}{2}MLML+2 \dot{L}M^2\big)\\
  \ &= \  {\rm tr}\big(M^2L^2+\tfrac{1}{2}MLML+2M^4\big)+2\,\dot{\Phi}\,{\rm tr}(M^2L)\\
  \ & \simeq  \  {\rm tr}\big(-\tfrac{1}{2}MLML+M^4\big)\;.    
 \end{split}
 \ee 
For the $H^4$ invariants we find 
 \be
 H^4 
  \ = \ 3\,{\rm tr}(M^4)\,, \qquad 
 (H^2_{\mu\nu})^2  \ = \ 
  4\, {\rm tr}(M^4)+({\rm tr}(M^2))^2\,, \qquad
   (H^2)^2 \ = \   9  \, (\hbox{tr} \, M^2 )^2  \,. 
 \ee 
We need to calculate the terms with dilaton derivatives.  
Here we can use the lapse equation (as well as the other equations) to replace
\be
\dot \Phi^2 \ \to \  \tfrac{1}{4} \, \hbox{tr} (L^2-M^2)\;, 
\ee 
and thus simplify some of the calculations. 
First, one finds 
\be
\begin{split}
H^2_{\mu\nu} \, \partial^\mu\phi  \partial^\nu\phi \ = \ &
-\hbox{tr} (M^2) \, \dot \phi^2 \ = \ -\tfrac{1}{4} 
\hbox{tr} (M^2) \, \bigl( \dot \Phi + \tfrac{1}{2} \hbox{tr} L \bigr)^2 \\[1.0ex]
\ = \ & \   -\tfrac{1}{4} \dot \Phi^2  \hbox{tr} (M^2) \, 
-\tfrac{1}{4} \dot \Phi \,  \hbox{tr} (M^2) \hbox{tr} (L) 
\, -\tfrac{1}{16}  \,\hbox{tr} (M^2) (\hbox{tr} \, L)^2 \\[1.0ex]
\ \simeq  \ &   
-\tfrac{1}{16} \, \hbox{tr} (L^2-M^2)\,   \hbox{tr} (M^2) \, 
+\tfrac{1}{4} \hbox{tr} (M^2 L)   \hbox{tr} (L) \, + \tfrac{1}{8} (\hbox{tr} M^2)^2 
\, -\tfrac{1}{16}  \,\hbox{tr} (M^2) (\hbox{tr} \, L)^2 \\[1.0ex]
\ \simeq  \ &   
-\tfrac{1}{16} \,\hbox{tr} (M^2)\, \hbox{tr} (L^2) 
\, + \tfrac{3} {16} (\hbox{tr} M^2)^2\, 
+\tfrac{1}{4} \hbox{tr} (M^2 L)   \hbox{tr} (L) \, 
\, -\tfrac{1}{16}  \,\hbox{tr} (M^2) (\hbox{tr} \, L)^2\;. 
 \end{split}
\ee 
The next structure, 
reduced to one-dimension, is  
proportional to the previous one 
\be
H^2 (\partial \phi)^2 \ = \ - 3 \, \hbox{tr} (M^2) \, \dot\phi^2 \ = \ 3 
H^2_{\mu\nu} \, \partial^\mu\phi  \partial^\nu\phi \,. 
\ee 
Therefore,   
\be
\gamma_6 \,  H^2_{\mu\nu} \, \partial^\mu\phi  \partial^\nu\phi 
  \ +\ \gamma_7\,  H^2 (\partial \phi)^2  \ = \ (\gamma_6 + 3 \gamma_7) 
 H^2_{\mu\nu} \, \partial^\mu\phi  \partial^\nu\phi \ \equiv \
 \tilde \gamma_6  H^2_{\mu\nu} \, \partial^\mu\phi  \partial^\nu\phi\,, 
\ee 
 and we have put here the new constant $\tilde \gamma_6$.  The
 condition of duality 
 invariance in one dimension 
 cannot completely determine the action.
 The last term takes a bit of effort,  
 giving 
 \be
 \begin{split}
 (\partial \phi)^4 \ = \ & \  \dot \phi^4 \ = \ \tfrac{1}{16} \bigl( 
  \dot \Phi + \tfrac{1}{2} \hbox{tr} L \bigr)^4 \\[1.0ex]
  \ = \ &  \ \tfrac{1}{16} \, \dot\Phi^4  \, + \tfrac{1}{8} \dot\Phi^3 \, 
  \hbox{tr}\, L  \, + \tfrac{3}{32} \dot\Phi^2 (\hbox{tr} \, L)^2  \, 
  + \tfrac{1}{32} \, \dot\Phi \, (\hbox{tr} \, L)^3  \, 
  + \tfrac{1}{(16)^2} \, (\hbox{tr}\, L)^4 \\[1.0ex]
  \ \simeq  \ &  \  \tfrac{1}{16}  \bigl( \, \hbox{tr} (L^2 - M^2) \bigr)^2  
   \,   -  \tfrac{1}{64}  \, \hbox{tr} (L^2 - M^2) \, \hbox{tr} (M^2)   
      + \tfrac{3}{128}  \, \hbox{tr} (L^2 - M^2) \, (\hbox{tr}\,L)^2    \\[1.0ex]
     &  - \tfrac{3}{64}  \, \hbox{tr} (M^2) \, (\hbox{tr}\,L)^2    
       + \tfrac{1}{(16)^2} \, (\hbox{tr}\, L)^4 \,.
       \end{split}
 \ee
We note here the presence of a term $(\hbox{tr}\, L)^4$.  No other contribution
to the left hand side of (\ref{vm-vm!}) contains such term,  
nor is it contained
on the right-hand side, as shown in (\ref{rhs-vm}).  This means that
$(\partial \phi)^4$ is incompatible with duality and we can immediately
set
\be
\gamma_8 \ = \ 0 \,. 
\ee 
The complete evaluation of the left hand side of (\ref{vm-vm!}) thus gives 
\be
\begin{split}
{\rm lhs} \ \simeq \ \  & \ \ \gamma_1 \Bigl(     
 \hbox{tr} \bigl(   \tfrac{1}{8} L^4\, 
 -MLML\, -\tfrac{3}{2} M^2 L^2 \bigr)   +  \tfrac{3}{8}  \bigl( \hbox{tr}\, L^2 
 \bigr)^2\,  - 
 \tfrac{1}{4} \,\hbox{tr} M^2 \, \hbox{tr} L^2 \Bigr) \\[1.0ex]
 & + \gamma_2 \Bigl(  {\rm tr}\big(-\tfrac{1}{2}MLML+M^4\big) \Bigr) 
 \  + \gamma_3 \Bigl( 3 \hbox{tr} (M^4) \Bigr) \\[1.0ex]
 &  + \gamma_4 \Bigl(  4\, {\rm tr}(M^4)+({\rm tr}(M^2))^2 \Bigr) 
 \, + \gamma_5 \Bigl(  9 (\hbox{tr} M^2)^2\Bigr) \\[1.0ex]
  & + \tilde 
  \gamma_6 \Bigl(
  -\tfrac{1}{16} \,\hbox{tr} (M^2)\, \hbox{tr} (L^2) 
\, + \tfrac{3} {16} (\hbox{tr} M^2)^2\, 
+\tfrac{1}{4} \hbox{tr} (M^2 L)   \hbox{tr} (L) \, 
\, -\tfrac{1}{16}  \,\hbox{tr} (M^2) (\hbox{tr} \, L)^2 \Bigr) \;. 
\end{split}
\ee
Equating the coefficients of the independent structures on the left- and right-hand
sides of (\ref{vm-vm!}) we get ten equations: 
\be
\begin{split}
\tfrac{1}{8} \gamma_1 \ = \ & \  2a_1 \,, \\[1.0ex]
- (\gamma_1 + \tfrac{1}{2} \gamma_2) \ = \ & \  4a_1 + \tfrac{2}{3} (b_1+ \tilde b_2 )  \,, \\[1.0ex]
- \tfrac{3}{2} \gamma_1  \ = \ & \ - 8a_1+ \tfrac{2}{3} (b_1+ \tilde b_2)\,, \\[1.0ex]
 \gamma_2 + 3\gamma_3  + 4 \gamma_4 \ = \ & \ 
2a_1 + \tfrac{2}{3} (b_1+ \tilde b_2 )\,, \\[2.0ex]
\tfrac{3}{8} \gamma_1   \ = \ & \  
4a_2 - \tfrac{1}{3} b_1 + \tfrac{1}{12} b_4 \,, \\[1.0ex]
  -\tfrac{1}{4} \gamma_1 -\tfrac{1}{16} \tilde\gamma_6 
 \ = \ & \ 
 - 8a_2  + \tfrac{1}{3} b_1 - \tfrac{1}{3} \tilde b_2 - \tfrac{1}{6} b_4 \,, \\[1.0ex]
  \gamma_4  + 9 \gamma_5 + \tfrac{3} {16}\tilde\gamma_6
      \ = \ & \  
 4a_2  + \tfrac{1}{3} \tilde b_2 + \tfrac{2}{3} b_3  + \tfrac{1}{12} b_4\,,  \\[2.0ex]
   \tfrac{1}{4} \tilde\gamma_6  \ = \ & \  \tfrac{4}{3}  b_3 \,,  \\[1.0ex]
- \tfrac{1}{16} \tilde\gamma_6 
\ = \ & \ \tfrac{1}{3} b_3 \,,    \\[1.0ex]
0    \ = \ &  - \tfrac{1}{3} b_3  \,.
\end{split}
\ee
The first four equations and the last three equations are completely
equivalent to the following  values
for the parameters  
\be
a_1 \ =  \  \tfrac{1}{16} \gamma_1 \,, \qquad
b_1 + \tilde b_2 \ = \   -\tfrac{3}{2} \gamma_1\,, \qquad
b_3 \ = \  0 \,, 
\ee
as well as the following constraints on the action coefficients:
\be
\gamma_2  \ = \  -\tfrac{1}{2} \gamma_1 \,, \qquad 
3\gamma_3 + 4 \gamma_4  \ = \  -\tfrac{3}{8} \gamma_1\,, \qquad 
\tilde \gamma_6 \ = \  0\,. 
\ee
With this information, the remaining equations 
are
\be
\begin{split}
\tfrac{3}{8} \, \gamma_1    \ = \ & \  
4a_2 - \tfrac{1}{3} b_1 + \tfrac{1}{12} b_4\,,  \\[1.0ex]
  -\tfrac{1}{4} \gamma_1   \ = \ & \ 
 - 8a_2  + \tfrac{1}{3} b_1 - \tfrac{1}{3} \tilde b_2 - \tfrac{1}{6} b_4  \,,
 \\[1.0ex]
  \gamma_4  + 9 \gamma_5    \ = \ & \  
 4a_2  + \tfrac{1}{3} \tilde b_2   + \tfrac{1}{12} b_4  \,.
  \end{split}
\ee
Interestingly, if we add the three of them, the right-hand side vanishes and
we get one more constraint for the action coefficients:
\be
 \gamma_4  + 9 \gamma_5 \ = \ -\tfrac{1}{8} \, \gamma_1 \, . 
\ee
There are no more constraints from these equations and 
the full list of constraints is then
\be
\gamma_2  \ =  \  -\tfrac{1}{2} \gamma_1 \,, 
\quad  3\gamma_3 + 4 \gamma_4  \ = \   -\tfrac{3}{8} \gamma_1 \,, \quad
 \gamma_4  + 9 \gamma_5 \ = \   -\tfrac{1}{8} \, \gamma_1\,, \quad
\tilde \gamma_6 \ = \ 
\gamma_8 \ = \   0\,.
\ee
We can parameterize the coefficients $\gamma_3, \gamma_4,$ and $\gamma_5$
 in terms
of a parameter $t$,  and $\gamma_6$ and $\gamma_7$ in terms of a parameter
$u$ (recalling that $\tilde \gamma_6= \gamma_6 + 3 \gamma_7 =0$).
We then get  
 \be
 \begin{split}  
{\cal L}(\gamma) \ = \  \gamma_1\, \Bigl( \ & 
  \, R_{\mu\nu\rho\sigma}^2  
 \  - \tfrac{1}{2}  \,  HHR 
 \  +\ \bigl( \tfrac{1}{24} + t \bigr)  \, H^4  
 \ +\ \bigl( - \tfrac{1}{8} - \tfrac{3}{4} t\bigr) \,  (H^2_{\mu\nu} )^2 \\[1.0ex]
&  \ +\  \tfrac{1}{12} t \,   (H^2)^2 
 \ +\ u \,  H^2_{\mu\nu} \, \partial^\mu\phi  \partial^\nu\phi 
  \ -\ \tfrac{1}{3} u \,  H^2 (\partial \phi)^2    \Bigr) \,. 
\end{split}
\ee
If we take $t=u =0$ we recover the known T-duality invariant 
action. 
Up to an overall constant, the eight
 coefficients in the action are really seven coefficients.
From these seven we have determined five, since we have two free parameters. 
The $u$ parameter dependence is such that  
reduced to one dimension it disappears.  
One can quickly check that the $t$ dependence also vanishes in a reduction
to one dimension. 
The constraint of T-duality in the 
reduction thus had no hope to determine 
$t$ nor $u$.

\sectiono{Green-Schwarz term and T-duality at ${\cal O}(\alpha'^2)$} 

We now turn to a different application.   
The goal it to investigate what the double field theory (DFT) 
constructed in \cite{Hohm:2013jaa} is in terms of conventional field variables. 
This theory, which we call DFT$^{-}$ \cite{Hohm:2014xsa}, 
features the deformed gauge transformations of the Green-Schwarz mechanism 
\cite{Green:1984sg}, 
as shown in \cite{Hohm:2014eba}.   
 In conventional language this  
implies that the field strength $H$ must be replaced by
the improved field strength $\widehat H$ that includes the Chern-Simons 
term built from the Christoffel connection. 
Thus, the minimal action   
consistent with the two-derivative theory and gauge invariance reads  
\be
\label{GSdeformedAction}
S \ = \   \int d^Dx  \sqrt{-g}\,  e^{-2\phi} \Bigl( R +  4 (\partial \phi)^2 -\tfrac{1}{12} \widehat H^2   \Bigr)\;, 
\ee
where
\be
\widehat H_{\mu\nu\rho} ( b, \Gamma)  \ =  \ 3\, \bigl( \, \partial_{[\mu} \, b_{\nu\rho]} 
\ + \ \alpha' \Omega (\Gamma)_{\mu\nu\rho} \bigr) \,, 
\ee
with the Chern-Simons three-form 
\be
\Omega (\Gamma)_{\mu\nu\rho} \ = \ \Gamma^\alpha_{[\mu|\beta | } \partial^{}_\nu
\Gamma^\beta_{\rho] \alpha}   \, + \, \tfrac{2}{3} \, 
\Gamma^\alpha_{[\mu | \beta |}  \Gamma^\beta_{\nu| \gamma| } 
\Gamma^\gamma_{\rho] \alpha}  \,.
\ee
The full   
DFT$^{-}$ action might  
contain  
order $\alpha'^2$ terms beyond those following from the Chern-Simons modification in the above minimal action, 
but this is not required by gauge invariance:  
the minimal action is exactly gauge-invariant under
the deformed gauge transformations of the Green-Schwarz
mechanism.  
The purpose of this section is to 
test duality invariance in order to 
decide if the above minimal action could be   
the complete action of DFT$^{-}$. 
We will
see that while 
this minimal action satisfies duality invariance
to order $\alpha'$, it 
does not satisfy duality invariance to order $\alpha'^2$.
Thus the above action is not 
duality complete,
and we conclude that DFT$^{-}$ contains further higher-derivative invariants, 
whose determination we leave for future work. 

In order to reduce the action to one dimension
one first verifies that the only
non-vanishing Chern-Simons components are
\be
3 \Omega(\Gamma)_{0ij} \ = \ -\tfrac{1}{2}  \, g^{kl} \dot g_{k[i} \, \ddot g_{j] l} \,. 
\ee
It then follows that the only non-vanishing component of 
$\widehat H$ is
\be
\widehat H_{0ij} \ = \ \dot b_{ij} -\tfrac{1}{2} \, \alpha'  \, g^{kl} \dot g_{k[i} \, \ddot g_{j] l}\,, 
\ee
and in matrix notation:
\be
\widehat H_0{}^i{}_j \ =  \ \bigl(  M    - \tfrac{1}{4} \alpha'  \, [ L , \dot L ] \bigr)^i{}_j \,. 
\ee
In order to simplify the evaluation of traces we often use the 
transposition properties
\be
L^T \ =  \ g L g^{-1}  \,, \quad 
M^T \ =  \ -g M g^{-1}  \,, \quad   \dot L^T  \ = \ g \dot L g^{-1} \,.
\ee
Together with $\hbox{tr} \, Q =  \hbox{tr}\,  Q^T$, these allow us to
show that 
 \be\label{LMtraceID}
  \hbox{tr} (M^{2k+1} L^p) \ = \ 0  
  \qquad \text{for any  integers $k, p \ \geq \ 0$}\;. 
 \ee 
A short computation then gives for the $\widehat H^2$ term in the 
Lagrangian
\be
\label{h-sqr}
- \tfrac{1}{12} \widehat H_{\mu\nu\rho}^2 \ = \ -\tfrac{1}{4} \,\hbox{tr} ( \widehat H \widehat H) \ = \ -\tfrac{1}{4} \hbox{tr} M^2 
-\tfrac{1}{4} \,\alpha' \hbox{tr} (M\dot L L)   
+\tfrac{1}{32} \alpha'^2 
\hbox{tr} (L^2 \dot L^2  - (L \dot L)^2) \,.  
\ee
The $\alpha'$ correction here is actually removable by a field redefinition, thus making
it clear that to ${\cal O}(\alpha')$ the test of T-duality invariance
works out. 
Indeed, by the replacement $\dot L \to M^2 + \dot \Phi L$ we have
\be
\hbox{tr} (M\dot L L)  \   \simeq \ \hbox{tr} (M^3 L)   + \dot \Phi\,  \hbox{tr} (ML^2)  \ = \ 0\,,
\ee
since these traces are zero.
Even though the $\alpha'$ correction is removable, 
for the following analysis we will keep this term, 
but it is convenient to rewrite it in terms of a metric
field equation.  
To do this we first note that the metric and $b$-field variation 
in the two-derivative reduced theory (\ref{vmsrllctij}) gives
\be
\label{varetaHsqr-vm}
\begin{split}
\delta\Bigl( -\tfrac{1}{8} e^{-\Phi}\,    \,\hbox{tr} \bigl[ (\eta \dot \H)^2 \bigr] \Bigr)\ = \ & \ 
  \tfrac{1}{2}  \, e^{-\Phi}  \, \hbox{tr} \, \bigl[ \ 
 {\delta b}\,  g^{-1} \,{\cal B} \,g^{-1}  
\, - \, \, g^{-1} \delta g \, {\cal G} g^{-1}   \bigr]\, ,
\end{split}  
\ee
where   ${\cal G}$ and ${\cal B}$ are 
 given by
\be\label{calGB}
\begin{split}
  g^{-1}{\cal G} \ \equiv \ & \ \dot{L}-M^2-\dot\Phi\, L\,, \\[0.5ex]
  g^{-1}{\cal B} \ \equiv \ & \ \dot{M}-ML -\dot\Phi\, M\, .
\end{split}
 \ee 
Using the identities (\ref{LMtraceID}),  
we thus have 
\be
\hbox{tr} (M\dot L L)  \   =  \ \hbox{tr} \bigl(\,
LM ( g^{-1}{\cal G} +M^2+\dot\Phi\, L ) \bigr) \ = \  \hbox{tr} \bigl(\,
LM \,  g^{-1}{\cal G}   \bigr)\;. 
\ee
It is convenient to symmetrize the factor multiplying ${\cal G}$,
which we do by adding the trace of the transposed matrix 
and dividing by two:
\be
\hbox{tr} (M\dot L L)  \   =    \  \tfrac{1}{2} \hbox{tr} \bigl(
\big[ L\,, M\big]  \,  g^{-1}{\cal G}   \bigr)\,.
\ee
Thus, the reduced action finally 
takes the form
\be
\label{cov-red-1dim}
\begin{split}
S_{\rm red}
\ = \ &  \int dt \, e^{-\Phi}  \Bigl(  -\, \dot \Phi^{\,2} - \tfrac{1}{8}  \hbox{tr} (\eta \dot\H)^2 
 - \tfrac{1}{8} \alpha'   \,\hbox{tr} \bigl(  \big[L, M\big]  g^{-1} {\cal G}  \bigr)  
 + \tfrac{1}{32} \,\alpha'^2 \,  {\rm tr}\big(L^2\dot{L}^2-(L\dot{L})^2\big)    \,\Bigr) \,. 
\end{split}
\ee

\medskip 
We are now ready to test the   
duality invariance of the above  
action. To this end 
we consider 
the most general 
duality covariant action to order 
$(\alpha')^2$,   
\be\label{dualACTION}
S_{\rm dual} \ = \   \int dt \, e^{-\Phi}  \Big(  - \, \dot \Phi^{\,2} - \tfrac{1}{8}  \hbox{tr} (\eta \dot\H)^2  + \alpha'^2  {\cal L}^{(2)}  \Bigr) \,, 
\ee
where, given 
the general classification in (\ref{six-der}), we have 
\be
{\cal L}^{(2)}  \ = \    \  
a_1\,{\rm tr}(\eta\dot{\cal H})^6+a_2\,{\rm tr}(\eta\dot{\cal H})^4\,{\rm tr}(\eta\dot{\cal H})^2
+a_3\big({\rm tr}(\eta\dot{\cal H})^2\big)^3\,.
\ee
We  have to allow for field redefinitions in $S_{\rm dual}$,
implemented by the following replacements  
inside the generalized metric ${\cal H}(g,b)$: 
\be
\label{full-replace}
\begin{split}
g\  \to & \ \ g \, + \,  \alpha' \delta^{(1)} g  + \alpha'^2  \delta^{(2)} g\,, 
\qquad  \Delta \ \equiv \  \, g^{-1} 
\delta^{(1)}g  \;, 
\\[0.5ex]
b \ \to & \ \ b  \, \, + \,  \alpha' \cdot 0 \ \   + \alpha'^2  \delta^{(2)} b\,. 
\end{split}
\ee
The goal is to choose these redefinitions in such a way that we obtain
the dimensionally reduced 
action  (\ref{cov-red-1dim}).
Note that to first order in $\alpha'$ we only redefine the metric, but
to second order in $\alpha'$ both the metric and the antisymmetric
tensor are redefined.   
In order to compute the effect of the $\delta^{(2)}$ redefinitions to order 
$\alpha'^2$
we just need the first variation of the two-derivative term $\hbox{tr} (\eta \dot {\cal H})^2$
given in (\ref{varetaHsqr-vm}). 
To compute the effect of the $\delta^{(1)}g$ redefinition to 
order 
$\alpha'^2$  
we need the {\em second}   
 variation of the two-derivative term under a change of the metric. 
Denoting 
this change of the metric
by 
$\Delta$, 
\be
   g \ \to   \ g + \delta g  \,, \quad   \quad   \Delta \ \equiv \  g^{-1} \delta g\,, 
\ee
a calculation gives
\be
\label{sec-var-vmvm}
\begin{split}
 - \tfrac{1}{8}\, e^{-\Phi}\,
  \hbox{tr} (\eta \dot \H)^2\bigl|_{g+ \delta g}   \ =  \ 
  - \tfrac{1}{8}\, e^{-\Phi}\,\hbox{tr} (\eta \dot \H)^2\bigl|_{g} \ 
  & -    \,    \tfrac{1}{4}\, e^{-\Phi} \, \hbox{tr} \bigl( -2 \Delta 
  M^2  -2 L \dot \Delta \bigr)  \\[0.8ex]
  &    - \, \tfrac{1}{4}\, e^{-\Phi} \, \hbox{tr} \bigl( 
    \Delta M \Delta M + 2 \Delta^2 M^2 
  -  \dot \Delta \dot \Delta  + 2\dot \Delta \Delta L \bigr) \,.  
\end{split}
\ee
The second order variation is on the second line of the right-hand 
side.

We now perform  
the replacement (\ref{full-replace}) 
in the duality covariant action (\ref{dualACTION}).
Using the general first variation (\ref{varetaHsqr-vm}),  
the second variation from (\ref{sec-var-vmvm}), and letting
$\Delta \equiv  \, g^{-1} \delta^{(1)}g$  we find  
\be\label{replacedaction}
\begin{split}
S_{\rm dual}\bigl|_{\rm rep} \ = \ &  \int dt \, e^{-\Phi}  \Bigl(  -\, \dot \Phi^{\,2} - \tfrac{1}{8}  \hbox{tr} (\eta \dot\H)^2 \,   
- \tfrac{1}{2}  \, \alpha'  \, \hbox{tr} \, \bigl[ \, \Delta\,  g^{-1} {\cal G}  \bigr]\\[1ex]
&   + \tfrac{1}{2}  \, \alpha'^2  \, \hbox{tr} \, \bigl[ \ 
{\delta^{(2)} b} \,g^{-1}{\cal B}g^{-1}    
\,- \, \delta^{(2)} g \,  g^{-1} {\cal G}g^{-1}    \bigr] \\[0.8ex]
& - \,\tfrac{1}{4} \,\alpha'^2\,  \hbox{tr} \,\bigl(  \Delta M 
\Delta M + 2 \Delta^2 M^2 
  -  \dot \Delta \dot \Delta  + 2\dot \Delta \Delta L  \bigr) \ +\alpha'^2 {\cal L}^{(2)}  + {\cal O} (\alpha'^3) \Bigr)  \, . 
\end{split}
\ee
Comparing with the dimensionally reduced action
(\ref{cov-red-1dim}), we infer that we need to choose 
\be
\Delta   
\ = \ \tfrac{1}{4} \big[ L, M\big] \,, 
\ee
in order to match it to first order in $\alpha'$.  Note that this 
is not a duality covariant field redefinition.
For this choice we find
for the second-order  variation 
\be
\begin{split}
- \tfrac{1}{4} \,  \hbox{tr} \bigl(  \Delta M \Delta M 
+ & \  2 \Delta^2 M^2 
-    \, \dot \Delta \dot \Delta  
+ 2\dot \Delta \Delta L \bigr)    \\[1.0ex]
& \  \simeq \  \tfrac{1}{32}\,  \hbox{tr} \, \bigl( ML^3 ML 
- ML^2 ML^2 \ + \ 
M^4 L^2 - M^3 L ML \bigr) \\[0.8ex]
&\quad   + \tfrac{1}{8} \, \dot \Phi^2  \,\hbox{tr} \,( MLML - M^2 L^2)\,.  
\end{split}
\ee
 Thus, inserting this into (\ref{replacedaction}), we get 
\be
\begin{split}
S_{\rm dual}\bigl|_{\rm rep} \ = \ &  \int dt \, e^{-\Phi}  \Bigl(  - \, 
\dot \Phi^{\,2} - \tfrac{1}{8}  \hbox{tr} (\eta \dot\H)^2 \,    - \tfrac{1}{8} \alpha'   \,\hbox{tr} \bigl(  \, [L, M] \ g^{-1} {\cal G}  \,  \bigr)
\\[1ex]
&   + \tfrac{1}{2}  \, \alpha'^2  \, \hbox{tr} \, \bigl[ \ 
{\delta^{(2)} b} \,g^{-1}{\cal B}g^{-1}    
\,- \,  \delta^{(2)} g \,  g^{-1} {\cal G}g^{-1}    \bigr] \\[0.5ex]
& +  \tfrac{1}{32}\, \alpha'^2  \hbox{tr} \, 
\bigl( ML^3 ML - ML^2 ML^2  +  
M^4 L^2 - M^3 L ML \bigr) \\[0.8ex]
& \  +  \tfrac{1}{8} \,\alpha'^2 \, \dot \Phi^2  \,\hbox{tr} \,
( MLML - M^2 L^2)  + 
\alpha'^2 {\cal L}^{(2)}  + {\cal O} (\alpha'^3) \Bigr) \,. 
\end{split}
\ee
We now note that up to further redefinitions we can replace 
 \be
  L^2\dot{L}^2-(L\dot{L})^2 \ \simeq  \ M^4L^2-M^2LM^2L\;, 
 \ee
where it is easy to see that no dilaton terms are produced. 
Therefore we can add to the above
Lagrangian the term 
\be
  \tfrac{1}{32} \,\alpha'^2 \,  {\rm tr}\big(L^2\dot{L}^2-(L\dot{L})^2\big)
 - \tfrac{1}{32} \,\alpha'^2 \,  {\rm tr}\big(M^4L^2-M^2LM^2L\big)\;, 
\ee
by absorbing the field redefinitions into the still undetermined 
$\delta^{(2)}b$ and $\delta^{(2)} g$.   
This, together with the use of  
(\ref{lapseREDEf}) for the term on the last line 
of the Lagrangian, gives
\be\label{FinalCOv}
\begin{split}
S_{\rm dual}\bigl|_{\rm rep} \ = \ &  \int dt \, e^{-\Phi}  \Bigl(  - \, \dot \Phi^{\,2} - \tfrac{1}{8}  \hbox{tr} (\eta \dot\H)^2 \,    - \tfrac{1}{8} \alpha'   \,\hbox{tr} \bigl(  \, [L, M] \ g^{-1} {\cal G}  \,  \bigr) 
+ \tfrac{1}{32} \,\alpha'^2 \,  {\rm tr}\big(L^2\dot{L}^2-(L\dot{L})^2\big) 
\\[1ex]
&   + \tfrac{1}{2}  \, \alpha'^2  \, \hbox{tr} \, \bigl[ \ 
{\delta^{(2)} b} \,g^{-1}{\cal B}g^{-1}    
\,- \,  \delta^{(2)} g \,  g^{-1} {\cal G}g^{-1}    \bigr] \\[1ex]
& +  \tfrac{1}{32}\, \alpha'^2  \hbox{tr} \, 
\bigl( ML^3 ML - ML^2 ML^2  +  
M^2 LM^2 L - M^3 L ML \bigr)  \\[0.8ex]
&   +  \tfrac{1}{32} \,\alpha'^2 \, \hbox{tr} (L^2-M^2)   \,\hbox{tr} \,
( MLML - M^2 L^2)  + 
\alpha'^2 {\cal L}^{(2)}  + {\cal O} (\alpha'^3) \Bigr)\,.
\end{split}
\ee

Let us now compare this with the dimensionally reduced action 
(\ref{cov-red-1dim}), 
 \be
\begin{split}
S_{\rm red} 
\ = \ &  \int dt \, e^{-\Phi}  \Bigl(  -\, \dot \Phi^{\,2} - \tfrac{1}{8}  \hbox{tr} (\eta \dot\H)^2 
 - \tfrac{1}{8} \alpha'   \,\hbox{tr} \bigl(  \, [L, M]  g^{-1} {\cal G}  \bigr)  
 + \tfrac{1}{32} \,\alpha'^2 \,  {\rm tr}\big(L^2\dot{L}^2-(L\dot{L})^2\big)   \ \Bigr) \,. 
\end{split}
\ee
This coincides exactly with the first line of (\ref{FinalCOv}). 
Therefore, 
the hypothesis that the 
action is duality invariant requires
that we can choose the duality covariant terms in ${\cal L}^{(2)}$ so that 
the final two lines of (\ref{FinalCOv}) are zero up to field and lapse redefinitions,  
thereby determining in particular $\delta^{(2)}g$ and $\delta^{(2)}b$ in the second line. 
By the procedure explained in sec.~2  (see eqn.~(\ref{finalduality})) 
this requires 
that we can choose coefficients $a_1,\ldots , a_3$ 
and a function $X$ such that 
\be
\label{Oddcoonnf-897-vm}
 \begin{split}
 0 \ \simeq & \ \   \tfrac{1}{32}\,  \hbox{tr} \, 
 \bigl( ML^3 ML - ML^2 ML^2 \ + \ 
M^2LM^2 L - M^3 L ML \bigr) \\[1.0ex]
&  \, + \tfrac{1}{32} \,
 \hbox{tr} ( L^2 -M^2)\,\hbox{tr} \,( MLML - M^2 L^2) 
 \\[0.5ex]
&\, +a_1\,{\rm tr}(\eta\dot{\cal H})^6+a_2\,{\rm tr}(\eta\dot{\cal H})^4\,{\rm tr}(\eta\dot{\cal H})^2
+a_3\big({\rm tr}(\eta\dot{\cal H})^2\big)^3\, \\[1.0ex]
&  \, + \bigl( \dot \Phi^2 - \tfrac{1}{4} \hbox{tr} ( L^2 -M^2) \bigr)  
X \;. 
\end{split}
\ee
We now make the most general ansatz for the function $X$ to this order in derivatives by writing 
$X = Y+ \tilde X$, where 
\be
\label{most-general-Y}
Y \ = \  \alpha_1 \,  \hbox{tr} L^4  
+ \alpha_2 \, \hbox{tr} (L^2M^2) + \alpha_3 \, \hbox{tr} (MLML) 
+ \alpha_4 \,\hbox{tr} M^4 \,,
\ee
is the most general ansatz with a single trace, 
no dilaton derivatives, and an even number of $M$'s.  
The term $\tilde X$ then includes
dilaton derivatives and or multiple traces.
This implies that 
 it cannot contribute relevant
terms with single traces.\footnote{Single trace terms 
in (\ref{Oddcoonnf-897-vm})
can arise from $\tilde X$ terms 
 of the form  $\dot \Phi^k W$, where $W$ has a single trace.  
Inserted in (\ref{Oddcoonnf-897-vm}), the single-trace terms 
can only arise  from  $\dot \Phi^{k+2} W$.
The recursive relation (\ref{recursive-vm}) then
implies that the single-trace contribution arises from 
$\dot \Phi^2 W^{' \ldots '}$, where $W$ is primed-differentiated $k$ times.
The effect of this contribution amounts to additive changes
to the coefficients $\alpha_i$
 in  (\ref{most-general-Y}), since $Y$ includes all possible single
traces.  This is an irrelevant contribution (if one only cares about 
single traces, as we do) since the coefficients in $Y$
are already completely general.  Finally, one can quickly check that 
a $\dot \Phi^4$ in $\tilde X$ can only give multiple traces.}  

  The idea now
is to show that the single trace part of (\ref{Oddcoonnf-897-vm})
cannot be satisfied.   To extract the
single traces from the last line of (\ref{Oddcoonnf-897-vm}) we recall the identity 
(\ref{dilatonIDentities-99-vm}). Since all terms in $Y$ have 
$k=4$, the contribution to the
single trace from $\dot \Phi^2 \, 
Y $ is proportional to $Y''$
\be
\dot\Phi^2   
 \, Y  |_{\rm s.t.}  \ \simeq \ 
 {2 \over k (2k -1)} \, Y'' \ \simeq \ 
 \tfrac{1}{14} \, Y''\,.
\ee
We find 
 \be
  \begin{split} 
   Y'' \ = \ \,& {\rm tr}\big[\,2(2\alpha_1+\alpha_2+\alpha_3)MLML^3+(4\alpha_1+\alpha_2)M^2L^4
   +(\alpha_2+2\alpha_3)ML^2 ML^2\\[1ex]
   &\quad +(8\alpha_1+5\alpha_2+4\alpha_3+4\alpha_4)M^4 L^2
   +(4\alpha_1+3\alpha_2+2\alpha_3+4\alpha_4)M^2L M^2L\\[1ex]
   &\quad +2(3\alpha_2+4\alpha_3+4\alpha_4)M^3LML
   +2(\alpha_2+\alpha_3+2\alpha_4)M^6\,\big]\;. 
  \end{split}
 \ee  
Since nowhere here there is an $L^6$ we cannot 
get the
full $\hbox{tr} (\eta \dot \H)^6$ and therefore we
must have $a_1 =0$.  
Moreover, as $\hbox{tr} (\eta \dot \H)^6$ is the only duality invariant single trace term to this order in 
derivatives, 
this means that we
must cancel the full set of four single trace terms on the
first line of (\ref{Oddcoonnf-897-vm}).   
In order to see that no solution 
exists 
it suffices
to collect terms in $Y''$ proportional to $M^2$:  
\be
\begin{split} 
\dot\Phi\, Y\big|_{\rm s.t.}    \ = \  &\,  (4\alpha'_1 + 2\alpha'_2 + 2\alpha'_3) {\rm tr}(ML^3 ML)  
+ (\alpha'_2 + 2\alpha'_3) {\rm tr}(ML^2 ML^2)  \\[0.5ex]
 &\, +  (4\alpha'_1 + \alpha'_2){\rm tr}(M^2 L^4)   + {\cal O} (M^4)  \,,
\end{split}
\ee
where we defined $\alpha_i' = 
\tfrac{1}{14} \alpha_i$.
To cancel the single traces 
in (\ref{Oddcoonnf-897-vm}) we
need
\be
4\alpha'_1 + 2\alpha'_2 + 2\alpha'_3\ = \ -\tfrac{1}{32} \,, \quad \alpha'_2 + 2\alpha'_3\ = \ \tfrac{1}{32} \,, \quad 4\alpha'_1 + \alpha'_2 \ = \ 0 \,. 
\ee
The use of the third equation means that the first and second equations
become, respectively,
\be
\alpha'_2 + 2\alpha'_3\ = \ -\tfrac{1}{32} \,, \quad \alpha'_2 + 2\alpha'_3\ = \ \tfrac{1}{32} \,,  
\ee
which has  no solution. 
 This proves that duality does not hold to ${\cal O}(\alpha'^2)$
for the action in 
(\ref{GSdeformedAction}).

\sectiono{Conclusions}

We have improved on a method by Meissner to test T-duality invariance
of actions with $\alpha'$ corrections.  The method 
is now systematic enough
that it can 
be used to test duality covariance to ${\cal O}(\alpha'^2)$. It 
works with an arbitrary field basis, 
so the analysis can begin with the simplest form of the action that can be obtained
by covariant field redefinitions.   We have emphasized a built-in limitation of the method: there are non-trivial linear combinations of terms that 
give zero upon dimensional reduction to one dimension.  Such linear combinations of terms cannot
be constrained by this test.   

The above test  
 of T-duality invariance of $\alpha'$ corrections does not suffice to prove
that a given action is duality invariant to a given order in $\alpha'$.  
It provides
a necessary but not sufficient condition for duality invariance. 
For the bosonic
or heterotic string a direct  
proof of T-duality invariance to order $\alpha'$ would be furnished
by an extension of the Maharana-Schwarz analysis~\cite{Maharana:1992my} 
to order $\alpha'$.  
An exactly duality-invariant effective action 
almost surely
would require terms of all orders in $\alpha'$.    While the test of T-duality
can be applied to string theories in the critical dimension, it could also be
applied to the low-energy limits and derivative corrections 
that arise after arbitrary compactifications,
as long as there remains spatial dimensions so that continuous T-duality would
emerge upon further compactification on tori.    

The power of a double field
theory formulation is that it proves T-duality just by its existence. 
The doubled $\alpha'$ geometry of~\cite{Hohm:2013jaa} 
furnishes an exactly T-duality invariant action with $\alpha'$ corrections.
This theory, called DFT$^{-}$, is a duality invariant completion of the Green-Schwarz mechanism. 
We have at present little idea how this action
looks in terms of conventional field variables 
beyond first order in $\alpha'$. 
We applied our test to learn that the action in conventional variables must have terms beyond those that arise from the 
minimal 
Green-Schwarz 
modification of the $b$-field field strength in the kinetic terms.

\section*{Acknowledgments} 
We thank Ashoke Sen and Krzysztof Meissner
for useful discussions on tests of T-duality. 
 The work of O.H. is supported by a DFG Heisenberg fellowship. 
The work of B.Z. is supported by the 
U.S. Department of Energy (DoE) under the cooperative 
research agreement DE-FG02-05ER41360.


\begin{thebibliography}{99}

%\cite{Giveon:1994fu}
\bibitem{Giveon:1994fu} 
  A.~Giveon, M.~Porrati and E.~Rabinovici,
  ``Target space duality in string theory,''
  Phys.\ Rept.\  {\bf 244}, 77 (1994)
  [hep-th/9401139].
  %%CITATION = HEP-TH/9401139;%%
  %788 citations counted in INSPIRE as of 26 Aug 2015
  
  
\bibitem{Veneziano:1991ek}   
  G.~Veneziano,
  ``Scale factor duality for classical and quantum strings,''
  Phys.\ Lett.\ B {\bf 265}, 287 (1991). \
  %%CITATION = PHLTA,B265,287;%%
  %628 citations counted in INSPIRE as of 17 Oct 2014
   %\cite{Meissner:1991zj}
%\bibitem{Meissner:1991zj} 
  K.~A.~Meissner and G.~Veneziano,
  ``Symmetries of cosmological superstring vacua,''
  Phys.\ Lett.\ B {\bf 267}, 33 (1991).
  %%CITATION = PHLTA,B267,33;%%
  %295 citations counted in INSPIRE as of 17 Oct 2014

%\cite{Sen:1991zi}
\bibitem{Sen:1991zi} 
  A.~Sen,
  ``O(d) x O(d) symmetry of the space of cosmological solutions in string theory, scale factor duality and two-dimensional black holes,''
  Phys.\ Lett.\ B {\bf 271}, 295 (1991).
  %%CITATION = PHLTA,B271,295;%%
  %174 citations counted in INSPIRE as of 17 Oct 2014

\bibitem{9109038} 
  S.~F.~Hassan and A.~Sen,
  ``Twisting classical solutions in heterotic string theory,''
  Nucl.\ Phys.\ B {\bf 375}, 103 (1992)
  [hep-th/9109038].
  %%CITATION = HEP-TH/9109038;%%

\bibitem{Maharana:1992my} 
  J.~Maharana and J.~H.~Schwarz,
  ``Noncompact symmetries in string theory,''
  Nucl.\ Phys.\ B {\bf 390}, 3 (1993)
  [hep-th/9207016].
  
  %\cite{Hohm:2014sxa}
\bibitem{Hohm:2014sxa} 
  O.~Hohm, A.~Sen and B.~Zwiebach,
  ``Heterotic Effective Action and Duality Symmetries Revisited,''
  JHEP {\bf 1502}, 079 (2015)
  [arXiv:1411.5696 [hep-th]].
  %%CITATION = ARXIV:1411.5696;%%
  %5 citations counted in INSPIRE as of 26 Aug 2015
  
   %\cite{Siegel:1993th,Hull:2009mi,Hohm:2010jy,Hohm:2010pp}
\bibitem{Siegel:1993th}
  W.~Siegel,
  ``Superspace duality in low-energy superstrings,''
  Phys.\ Rev.\ D {\bf 48} (1993) 2826
  [hep-th/9305073].
  
  %\cite{Siegel:1993bj}
\bibitem{Siegel:1993bj} 
  W.~Siegel,
  ``Manifest duality in low-energy superstrings,''
  In *Berkeley 1993, Proceedings, Strings '93* 353-363, and State U. New York Stony Brook - ITP-SB-93-050 (93,rec.Sep.) 11 p. (315661)
  [hep-th/9308133].  



%\cite{Hull:2009mi}
\bibitem{Hull:2009mi} 
  C.~Hull and B.~Zwiebach,
  ``Double Field Theory,''
  JHEP {\bf 0909}, 099 (2009)
  [arXiv:0904.4664 [hep-th]].
  %%CITATION = ARXIV:0904.4664;%%  

%\cite{Hohm:2010jy}
\bibitem{Hohm:2010jy} 
  O.~Hohm, C.~Hull and B.~Zwiebach,
  ``Background independent action for double field theory,''
  JHEP {\bf 1007}, 016 (2010)
  [arXiv:1003.5027 [hep-th]].

%\cite{Hohm:2010pp}
\bibitem{Hohm:2010pp} 
  O.~Hohm, C.~Hull and B.~Zwiebach,
  ``Generalized metric formulation of double field theory,''
  JHEP {\bf 1008} (2010) 008
   [arXiv:1006.4823 [hep-th]]. 

   
  
 %\cite{Hohm:2013jaa}
\bibitem{Hohm:2013jaa} 
  O.~Hohm, W.~Siegel and B.~Zwiebach,
  ``Doubled $\alpha'$-geometry,''
  JHEP {\bf 1402}, 065 (2014)
  [arXiv:1306.2970 [hep-th]]. 
    
%\cite{Hohm:2014eba}
\bibitem{Hohm:2014eba} 
  O.~Hohm and B.~Zwiebach,
  ``Green-Schwarz mechanism and $\alpha'$-deformed Courant brackets,''
  JHEP {\bf 1501}, 012 (2015)
  [arXiv:1407.0708 [hep-th]].
    
 %\cite{Hohm:2014xsa}
\bibitem{Hohm:2014xsa} 
  O.~Hohm and B.~Zwiebach,
  ``Double field theory at order $\alpha'$,''
  JHEP {\bf 1411}, 075 (2014)
  [arXiv:1407.3803 [hep-th]].
  

%\cite{Marques:2015vua}
\bibitem{Marques:2015vua} 
  D.~Marques and C.~A.~Nunez,
  ``T-duality and $\alpha'$-corrections,''
  arXiv:1507.00652 [hep-th].
  
 
%\cite{Meissner:1996sa}
\bibitem{Meissner:1996sa} 
  K.~A.~Meissner,
  ``Symmetries of higher order string gravity actions,''
  Phys.\ Lett.\ B {\bf 392}, 298 (1997)
  [hep-th/9610131].
  %%CITATION = HEP-TH/9610131;%%
  %70 citations counted in INSPIRE as of 26 Aug 2015
  
  
  %\cite{Godazgar:2013bja}
\bibitem{Godazgar:2013bja} 
  H.~Godazgar and M.~Godazgar,
  ``Duality completion of higher derivative corrections,''
  JHEP {\bf 1309}, 140 (2013)
  [arXiv:1306.4918 [hep-th]].
  %%CITATION = ARXIV:1306.4918;%%
  %10 citations counted in INSPIRE as of 26 Aug 2015


%\cite{Metsaev:1987zx}
\bibitem{Metsaev:1987zx} 
  R.~R.~Metsaev and A.~A.~Tseytlin,
  ``Order alpha-prime (Two Loop) Equivalence of the String Equations of Motion and the Sigma Model Weyl Invariance Conditions: Dependence on the Dilaton and the Antisymmetric Tensor,''
  Nucl.\ Phys.\ B {\bf 293}, 385 (1987).
  %%CITATION = NUPHA,B293,385;%%
  %330 citations counted in INSPIRE as of 26 Aug 2015
  
%\cite{Hohm:2015mka}  %bz the true reference
\bibitem{Hohm:2015mka} 
  O.~Hohm and B.~Zwiebach,
  ``Double Metric, Generalized Metric and $\alpha'$-Geometry,''
  arXiv:1509.02930 [hep-th].
  %%CITATION = ARXIV:1509.02930;%%
  


  %\cite{Green:1984sg}
\bibitem{Green:1984sg} 
  M.~B.~Green and J.~H.~Schwarz,
  ``Anomaly Cancellation in Supersymmetric D=10 Gauge Theory and Superstring Theory,''
  Phys.\ Lett.\ B {\bf 149}, 117 (1984).
  %%CITATION = PHLTA,B149,117;%%
  %2357 citations counted in INSPIRE as of 20 sept. 2015



\end{thebibliography}
\end{document}